\documentclass[%twocolumn,%showpacs,preprintnumbers,
amsmath,%amssymb,aps,
prd,nofootinbib,floatfix,11pt,%preprint,
]{revtex4}

\usepackage{graphicx}% Include figure files
\usepackage{setspace}
\usepackage{bm}% bold math

\usepackage[dvips]{color}
\definecolor{Red}{cmyk}{0,1,1,0}
\definecolor{BrickRed}{cmyk}{0,0.89,0.94,0.28}
\definecolor{Blue}{cmyk}{1,1,0,0}
\definecolor{Green}{cmyk}{1,0,1,0}

\newcommand\beq{\begin{eqnarray}}
\newcommand\eeq{\end{eqnarray}}
\def\lsim{\mathrel{\rlap{\lower4pt\hbox{$\sim$}}
    \raise1pt\hbox{$<$}}}                % less than or approx. symbol
\def\gsim{\mathrel{\rlap{\lower4pt\hbox{$\sim$}}
    \raise1pt\hbox{$>$}}}            

\allowdisplaybreaks
\newcommand\boldA{{\bf A}}
\newcommand\boldI{{\bf I}}
\newcommand\boldE{{\bf E}}
\newcommand\boldF{{\bf F}}
\newcommand\boldG{{\bf G}}
\newcommand\boldH{{\bf H}}
\newcommand\boldT{{\bf T}}

\newcommand\intA{A}
\newcommand\intI{I}
\newcommand\intE{E}
\newcommand\intF{F}
\newcommand\Fbar{\overline{F}}
\newcommand\intG{G}
\newcommand\intH{H}

\newcommand\Ieps{I_{\epsilon}}
\newcommand\Aeps{A_{\epsilon}}
\newcommand\Aepstwo{A_{\epsilon^2}}

\newcommand\Itwo{{I}_2}
\newcommand\Ione{{I}_1}
\newcommand\Izero{{I}_0}

\newcommand\Ethree{E_3}
\newcommand\Etwo{E_2}
\newcommand\Eone{E_1}
\newcommand\Ezero{E_0}

\newcommand\Fthree{F_3}
\newcommand\Ftwo{F_2}
\newcommand\Fone{F_1}
\newcommand\Fzero{F_0}

\newcommand\Gthree{G_3}
\newcommand\Gtwo{G_2}
\newcommand\Gone{G_1}
\newcommand\Gzero{G_0}

\newcommand\Hone{H_1}
\newcommand\Hzero{H_0}

\newcommand\lnbar{\overline{\ln}}
\newcommand\Lstwo{{\rm Ls_2}}
\newcommand\Lsthree{{\rm Ls_3}}
\newcommand\Lspfour{{\rm Ls'_4}}
\newcommand\zetathree{{\zeta_3}}

\newcommand\dilog{{\rm Li}_2}
\newcommand\trilog{{\rm Li}_3}

\newcommand\MSbar{$\overline{\rm{MS}}$ }

\begin{document}

\renewcommand{\theequation}{\arabic{section}.\arabic{equation}}
\renewcommand{\thefigure}{\arabic{section}.\arabic{figure}}
\renewcommand{\thetable}{\arabic{section}.\arabic{table}}

\title{\Large \baselineskip=20pt 
Evaluation of the general 3-loop vacuum Feynman integral}

\author{Stephen P.~Martin$^{1}$ and David G.~Robertson$^2$}
\affiliation{
\mbox{\it $^1$Department of Physics, Northern Illinois University, DeKalb IL 60115}\\
\mbox{\it $^2$Department of Physics, Otterbein University, Westerville OH 43081}\\
}

\begin{abstract}\normalsize \baselineskip=17pt 
We discuss the systematic evaluation of 3-loop vacuum integrals with 
arbitrary masses. Using integration by parts, the general integral of this 
type can be reduced algebraically to a few basis integrals. We define a set of
modified finite basis integrals that are particularly convenient for
expressing renormalized quantities. The basis 
integrals can be computed numerically by solving coupled 
first-order differential equations, using as boundary conditions 
the analytically known special cases that depend on only one mass scale. 
We provide the results necessary to carry this out, and introduce an 
implementation in the form of a public software package called 
{\tt 3VIL} (3-loop Vacuum Integral Library), 
which efficiently computes the numerical values of
the basis integrals for any specified masses.
{\tt 3VIL} is written in C, and can be linked from C, C++, or FORTRAN code. 
\end{abstract}

\maketitle

\vspace{-0.2in}

\tableofcontents

\baselineskip=15.4pt

%%%%%%%%%%%%%%%%%%%%%%%%%%%%%%%%%%%%%%%%%%%%%%%%%%%%%%%%%%%%%%%
\section{Introduction \label{sec:intro}}
\setcounter{equation}{0}
\setcounter{figure}{0}
\setcounter{table}{0}
\setcounter{footnote}{1}

With the discovery of the Higgs boson, the Standard Model has reached a 
milestone of experimental completion. Because all of the particle masses 
and couplings are now known directly or indirectly with well-defined 
experimental precisions, it is worthwhile to extend the calculations of 
the predictions of the Standard Model, as well as competitor extensions 
of it, to the kind of accuracy that requires loop integrals to be 
calculated beyond 2-loop order. In general it is useful to reduce 
theoretical uncertainties to the level at which they are completely 
negligible compared to the corresponding experimental and parametric 
errors. In some cases, the only reliable way to obtain estimates of 
theoretical error of a given calculation is to compute to an additional 
order in perturbation theory. In this paper, we address the problem of 
calculating the general 3-loop vacuum Feynman integral in dimensional 
regularization, with arbitrary propagator 
masses.\footnote{A numerical solution of general 3-loop 
vacuum integrals has also independently been obtained by A.~Freitas \cite{Freitas} 
in a way different from ours, namely in terms of
1-dimensional (or, in the 6-propagator case, 2-dimensional) integral representations,
by making use of dispersion relations.}

The computation of 2-loop vacuum integrals with arbitrary masses has 
been reduced to polylogarithms or equivalent functions, see 
refs.~\cite{Ford:1992pn,Davydychev:1992mt,Davydychev:1995mq,Caffo:1998du,Espinosa:2000df}. 
At 3-loop order, the vacuum integrals with one non-zero mass have been 
solved \cite{Broadhurst:1991fi, Avdeev:1994db, Fleischer:1994dc, 
Avdeev:1995eu, Broadhurst:1998rz, Fleischer:1999mp, Schroder:2005va}. In 
some special cases, 3-loop integrals with two distinct non-zero 
masses are also known analytically 
\cite{Davydychev:2003mv,Kalmykov:2005hb,Kalmykov:2006pu,
Bytev:2009mn,Bekavac:2009gz,Bytev:2009kb,Bytev:2011ks,Grigo:2012ji}. 
These results are 
reviewed below in section \ref{sec:analytic}, in our notations, and a 
few new two-scale special cases are added. The program {\tt MATAD} 
\cite{MATAD} is available for computations of vacuum diagrams with one 
non-zero mass scale, and more generally can be used in conjunction with 
expansions in ratios of squared masses and external momenta (see, for 
example, \cite{Smirnov:2002pj,Kant:2010tf}). 

One obvious application of the results given below is to the 3-loop 
effective potential of a general theory, with the Standard Model and its 
supersymmetric extensions as particular cases. In the latter case it is 
not clear {\it a priori} what the ordering or hierarchies of the masses 
will turn out to be. Even in the Standard Model case, it is helpful to 
be able to perform and present
calculations in a way that does not require expansions in particular mass 
hierarchies. Therefore, in 
the following we use an approach that does not depend on such 
expansions. The evaluation of the integrals is performed using the 
differential equations method \cite{Kotikov:1990kg,Remiddi:1997ny}, 
\cite{Caffo:1998du}, \cite{Caffo:1998yd,Gehrmann:1999as,Caffo:2002ch, 
Caffo:2002wm,Caffo:2003ma,Martin:2003qz,TSIL,Henn:2013pwa,Henn:2013nsa,Ablinger:2015tua,Remiddi:2016gno}. Here we use expressions 
for the derivatives of the basis integrals with respect to their squared 
mass arguments. The method is implemented in a public open-source computer package 
called {\tt 3VIL} (3-loop Vacuum Integral Library), which is 
structurally similar to, and compatible with, our earlier program {\tt 
TSIL} \cite{TSIL} for the calculation of 2-loop self-energy basis 
integrals.

The rest of this paper is organized as follows. In section 
\ref{sec:basis}, we establish our notations and conventions, including 
the basis of 3-loop vacuum integrals to which all others can be reduced 
using the method of integration by parts \cite{Chetyrkin:1981qh}. Two 
related but distinct alternative sets of basis integrals are defined, 
one incorporating counterterms in such a way that renormalized 
quantities are efficiently written in terms of them. The relation 
between the $\epsilon$ expansions of the two sets of basis integrals is 
given in a section \ref{sec:epsexp}. The derivatives of the basis integrals with 
respect to the propagator squared masses and the renormalization scale 
are given in section \ref{sec:massderivs}. In section \ref{sec:analytic}, we 
review the known analytical special cases, all of which have only one or two distinct 
non-zero masses. Section \ref{sec:diffeq} provides the differential 
equations used to compute the 3-loop vacuum integrals in the more 
general case of arbitrary masses. Section \ref{sec:3VIL} describes the
implementation of these results and an introduction to our 
public and open-source computer program {\tt 3VIL}. Section 
\ref{sec:Outlook} contains some concluding remarks.

%%%%%%%%%%%%%%%%%%%%%%%%%%%%%%%%%%%%%%%%%%%%%%%%%%%%%%%%%%%%%%%
\section{Basis integral definitions \label{sec:basis}}
\setcounter{equation}{0}
\setcounter{figure}{0}
\setcounter{table}{0}
\setcounter{footnote}{1}

\begin{figure}[!bt]
\begin{minipage}[]{0.49\linewidth}
    \includegraphics[width=0.7\linewidth,angle=0]{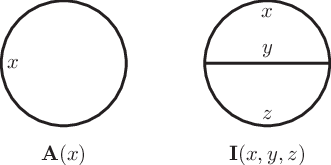}
\end{minipage}
\begin{minipage}[]{0.5\linewidth}
\caption{\label{fig:AI}
The topologies for the 1-loop  and 2-loop basis integrals 
$\boldA(x)$ and $\boldI(x,y,z)$
defined in eqs.~(\ref{eq:defA}) and (\ref{eq:defI}).}
\end{minipage}
\end{figure}

In this section, we establish our notational conventions and define 
the vacuum basis integrals up to 3-loop order. After Wick rotation, 
loop momentum integrations are carried out in
\beq
d = 4 - 2 \epsilon
\eeq
Euclidean dimensions, and denoted by
\beq
\int_p &\equiv& \mu^{4-d} \int \frac{d^dp}{(2\pi)^d} ,
\eeq
where $\mu$ is a regularization mass scale. 
Vacuum Feynman diagrams at 1-loop and 2-loop orders 
can be written in terms of the basis integrals 
$\boldA(x)$ and $\boldI(x,y,z)$ depicted in Figure \ref{fig:AI}.
%%%%%%%%%%%%%%%%%%%%%%%%%%%%%%%%%%%%%%%%%%%%%%%
%%%%%%%%%%%%%%%%%%%%%%%%%%%%%%%%%%%%%%%%%%%%%%%%%%%%%%%%%%%%%%%%%%
Here
\beq
\boldA(x) &=& 16 \pi^2 \int_p \frac{1}{p^2 + x} \,=\, 
\Gamma(-1+\epsilon)  \left (\frac{4\pi \mu^2}{x} \right )^{\epsilon} x ,
\label{eq:defA}
\eeq
where $x$ is the propagator squared mass. 
The two-loop order basis integral is defined by
\beq
\boldI(x,y,z) &=& (16 \pi^2)^2 \int_{p} \int_{q} 
\frac{1}{[p^2 + x][q^2 + y][(p-q)^2 + z]},
\label{eq:defI}
\eeq
which is symmetric on interchanges of any pair of squared masses $x,y,z$.
Any 2-loop vacuum Feynman diagram can be reduced to sums of $\boldI$
functions and products of two $\boldA$ functions, with coefficients that are ratios of
polynomials in the squared masses and the spacetime dimension $d$.

When presenting results for renormalized physical quantities 
(whether in the \MSbar scheme or any other scheme), it is convenient to 
eliminate the $\epsilon$-dependent
basis functions $\boldA$ and $\boldI$ 
in favor of $\epsilon$-independent functions
that include the effects of counterterms. 
We define
\beq
\lnbar(x) &\equiv& \ln(x/Q^2) ,
\eeq
with the \MSbar renormalization scale $Q$ defined by
\beq
Q^2 = 4 \pi e^{-\gamma_E} \mu^2 .
\label{eq:defQ2}
\eeq
Then we have the expansion:
\beq
\boldA(x) &=& -\frac{x}{\epsilon} + A(x) + \epsilon A_\epsilon(x) + 
\epsilon^2 A_{\epsilon^2}(x) + \ldots ,
\label{eq:defboldA}
\eeq
where\footnote{For brevity, we never include 
the common scale $Q$ explicitly among the arguments of loop integral 
functions.} 
\beq
A(x) &=& x [\lnbar(x)-1] ,
\\
\Aeps(x) &=& 
x\left [- \frac{1}{2} \lnbar^2(x) + \lnbar(x) -1 - \frac{\pi^2}{12} \right ] ,
\label{eq:defAeps}
\\
\Aepstwo(x) &=&
x\left [\frac{1}{6} \lnbar^3(x)- \frac{1}{2} \lnbar^2(x) 
+\left (1 + \frac{\pi^2}{12} \right ) \lnbar(x) 
- 1 - \frac{\pi^2}{12} + \frac{\zetathree}{3}\right ] .
\eeq

The $\epsilon$-expansion of the two-loop basis integral can be written as:
\beq
\boldI(x,y,z) &=& \frac{\Itwo(x,y,z)}{\epsilon^2}
+ \frac{\Ione(x,y,z)}{\epsilon}
+ \Izero(x,y,z)
+ \epsilon \Ieps(x,y,z)
+ \ldots,
\eeq
where the pole terms are 
\beq
\Itwo(x,y,z) &=& -(x+y+z)/2,
\label{eq:defItwo}
\\
\Ione(x,y,z) &=& \intA(x) + \intA(y) + \intA(z) -(x+y+z)/2.
\label{eq:defIone}
\eeq
However, 
instead of writing results in terms of $\Izero$ and $\Ieps$,
it is more convenient to follow\footnote{However, the notation is slightly different;
$I(x,y,z)$ in the present paper is equal to 
$(16 \pi^2)^2 \hat I(x,y,z)$ in ref.~\cite{Ford:1992pn}.} 
ref.~\cite{Ford:1992pn} 
by defining the ``renormalized" basis integral:
\beq
\intI(x,y,z) &=& \lim_{\epsilon \rightarrow 0} \left[
\boldI(x,y,z) 
- \intI^{(1)}_{\rm div}(x,y,z)
- \intI^{(2)}_{\rm div}(x,y,z) \right ] ,
\eeq
where the 1-loop and 2-loop ultraviolet (UV) sub-divergences are, respectively,
\beq
\intI^{(1)}_{\rm div}(x,y,z) &=& \frac{1}{\epsilon}[\boldA(x) + \boldA(y) + \boldA(z)] ,
\\
\intI^{(2)}_{\rm div}(x,y,z) &=& \frac{1}{2}(x+y+z)\left (\frac{1}{\epsilon^2} - \frac{1}{\epsilon}\right ) .
\eeq
Then one obtains:
\beq
\intI(x,y,z) &=& \Izero(x,y,z) - \Aeps(x) - \Aeps(y) - \Aeps(z).
\label{eq:relateII0}
\eeq
Now the 2-loop renormalized effective potential can be written efficiently 
in terms of $\intI(x,y,z)$ and $\intA(x)$,
as was done in ref.~\cite{Ford:1992pn} for the Standard Model and
ref.~\cite{Martin:2001vx} for general renormalizable theories,
without needing to use the functions $\Izero$  or $\Aeps$. 
For 3-loop renormalized quantities such as the effective potential, 
it is possible and natural to avoid the use of $\Ieps$, 
and the functions $\Izero$ and $\Aeps$ only appear 
in the combination $I$.
The integrals $\intI(x,y,z)$ and $\Izero(x,y,z)$ and $\Ieps(x,y,z)$ can be 
evaluated in terms of polylogarithms, using the methods of 
ref.~\cite{Ford:1992pn}. For completeness, these results 
are listed in section \ref{sec:analytic} below.

A general 3-loop order vacuum Feynman diagram will involve scalar 
integrals of the form shown in Figure \ref{fig:tetrahedron}:
%%%%%%%%%%%%%%%%%%%%%%%%%%%%%%%%%%%%%%%%%%%%%%%%%%%%%%%%%%%%%%%%
\begin{figure}[t]
\begin{minipage}[]{0.29\linewidth}
    \includegraphics[width=0.64\linewidth,angle=0]{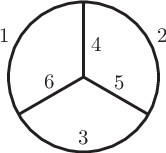}
\end{minipage}
\begin{minipage}[]{0.69\linewidth}
\caption{\label{fig:tetrahedron} The topology of the scalar Feynman diagram  
for the general 3-loop vacuum integral
$\boldT^{(n_1,n_2,n_3,n_4,n_5,n_6)}(x_1,x_2,x_3,x_4,x_5,x_6)$
defined in eq.~(\ref{eq:defT}).}
\end{minipage}
\end{figure}
%%%%%%%%%%%%%%%%%%%%%%%%%%%%%%%%%%%%%%%%%%%%%%%%%%%%%%%%%%%%%%%%
\beq
&& \boldT^{(n_1,n_2,n_3,n_4,n_5,n_6)}(x_1,x_2,x_3,x_4,x_5,x_6) \,=\,
(16 \pi^2)^3 \int_{p_1}\int_{p_2}\int_{p_3}
\nonumber
\\
&&
\frac{1}{[p_1^2 + x_1]^{n_1}
[p_2^2 + x_2]^{n_2}
[p_3^2 + x_3]^{n_3}
[(p_1-p_2)^2 + x_4]^{n_4} 
[(p_2-p_3)^2 + x_5]^{n_5}
[(p_3-p_1)^2 + x_6]^{n_6}}
,\phantom{xxxxxx}
\label{eq:defT}
\eeq
where the propagator powers $n_i$ can be positive, negative, or 0.
These integrals satisfy identities involving interchanges of
the pairs $(n_i, x_i)$, as implied by the tetrahedral
symmetry of the graphical representation shown in 
Figure \ref{fig:tetrahedron}. They also satisfy 9 identities implied by integration by
parts \cite{Chetyrkin:1981qh,Avdeev:1995eu,MATAD}:
\beq
0 &=& \int_{p_1}\int_{p_2}\int_{p_3} \frac{\partial}{\partial p_i^\mu}
\left [ p_j^\mu X \right ]
\eeq
for $i,j=1,2,3$, where $X$ is any product of propagators as in eq.~(\ref{eq:defT}).
The identities for $(i,j) = (1,1)$ and $(1,2)$ can be written as, acting on eq.~(\ref{eq:defT}),
\beq
0 &=&  d - 2 n_1 - n_4 - n_6 
+ 2 x_1 {\bf 1}^+ n_1 
+ (x_1 - x_2 + x_4 - {\bf 1}^- + {\bf 2}^- ) {\bf 4}^+ n_4 
\nonumber \\ && 
+ (x_1 - x_3 + x_6 - {\bf 1}^- + {\bf 3}^-) {\bf 6}^+ n_6 ,
\label{eq:trianglerule}
\\
0 &=& n_4 - n_1 
+ (x_1 + x_2 - x_4 - {\bf 2}^- + {\bf 4}^-) {\bf 1}^+ n_1 
+ (x_1 - x_2 - x_4 - {\bf 1}^- + {\bf 2}^-) {\bf 4}^+ n_4 
\nonumber \\ && 
+ (x_1 - x_3 - x_4 + x_5 - {\bf 1}^- + {\bf 3}^- + {\bf 4}^- - {\bf 5}^- ) {\bf 6}^+ n_6 ,
\eeq
and there are 2+5=7 other independent ones that can be
obtained from the above two as permutations implied by the tetrahedral symmetry. 
Here, the bold-faced raising and lowering operators 
are defined to increase or decrease the power of the corresponding propagator:
\beq
{\bf j}^\pm \boldT^{(\ldots,n_j,\ldots)}
= 
\boldT^{(\ldots,n_j\pm 1,\ldots)} .
\eeq
The dimensional analysis identity
\beq
0 &=& 3d/2 + \sum_{j=1}^6  (x_j {\bf j}^+ - 1) n_j.
\eeq
can also be obtained by combining the three integration by parts 
identities that involve $d$. Equation (\ref{eq:trianglerule}), and each of 11 identities
(not all independent) obtained by permutations of it using the symmetries of
the tetrahedron, is an example of what is sometimes known as the triangle rule.

\begin{figure}[t]
\includegraphics[width=0.92\linewidth,angle=0]{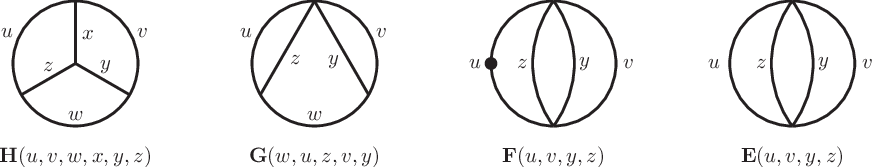}
\caption{\label{fig:HGFE}
The topologies for the integrals
$\boldH(u,v,w,x,y,z)$
and 
$\boldG(w,u,z,v,y)$
and
$\boldF(u,v,y,z)$
and
$\boldE(u,v,y,z)$
defined in eqs.~(\ref{eq:defH}), 
(\ref{eq:defG}),
(\ref{eq:defF}),
and (\ref{eq:defE}), respectively. 
The dot on the diagram for $\boldF(u,y,z,v)$ is used to 
indicate a squared propagator.
A complete basis for the 3-loop vacuum integrals
consists of integrals of the types $\boldH$, $\boldG$, $\boldF$, and
the products of 1-loop and 2-loop integrals $\boldA$ and $\boldI$.
The integral $\boldE$ is a useful adjunct, but is not included in the 
basis due to its redundancy, because of eq.~(\ref{eq:EEredundant}).}
\end{figure}

By repeated application of these integration by parts identities, any 3-loop vacuum 
integral $\boldT$ can eventually be reduced to a linear combination of integrals from a basis set, 
with coefficients that are ratios of polynomials in $d$ and the squared 
masses. The 
integrals in the basis are of five types, and can be defined as:
\beq
\boldH(u,v,w,x,y,z) &=& {\bf T}^{(1,1,1,1,1,1)}(u,v,w,x,y,z) ,
\label{eq:defH}
\\
\boldG(w,u,z,v,y) &=& {\bf T}^{(1,1,1,0,1,1)}(u,v,w,x,y,z),
\label{eq:defG}
\\
\boldF(u,v,y,z) &=& {\bf T}^{(2,1,0,0,1,1)}(u,v,w,x,y,z) ,
\label{eq:defF}
\\
\boldA(u) \boldI(v,w,y) &=& {\bf T}^{(1,1,1,0,1,0)}(u,v,w,x,y,z),
\\
\boldA(u) \boldA(v) \boldA(w) &=& {\bf T}^{(1,1,1,0,0,0)}(u,v,w,x,y,z) ,
\eeq
and the integrals obtained by 
permutations of the arguments of these according to the symmetries 
of the tetrahedron.
The last two are products of lower loop integrals, and 
therefore present no  problems. The integral defined by
\beq
\boldE(u,v,y,z) &=& {\bf T}^{(1,1,0,0,1,1)}(u,v,w,x,y,z)
\label{eq:defE}
\eeq
is useful, as we will see below, but it is 
not part of this canonical basis, because it can be reduced to the $\boldF$
integrals using the linear algebraic identity
\beq
\boldE(u,v,y,z) = \left [
u \boldF(u,v,y,z) +
v \boldF(v,u,y,z) +
y \boldF(y,u,v,z) +
z \boldF(z,u,v,y) 
\right ]/(-2 + 3 \epsilon) ,
\label{eq:EEredundant}
\eeq
which follows from dimensional analysis.
Note that
\beq
\boldF(u,v,y,z) = -\frac{\partial}{\partial u} \boldE(u,v,y,z) ,
\eeq
which allows some identities satisfied by the $\boldF$ integrals to be 
more easily derived or succinctly written in terms of the $\boldE$ 
integrals. The graph topologies associated with the functions $\boldH$, 
$\boldG$, $\boldF$, and $\boldE$ are shown in Figure \ref{fig:HGFE}.
%%%%%%%%%%%%%%%%%%%%%%%%%%%%%%%%%%%%%%%%%%%%%%%%%%%%%%%%%%%%%%%
%%%%%%%%%%%%%%%%%%%%%%%%%%%%%%%%%%%%%%%%%%%%%%%%%%%%%%%%%%%%%%%
We note that the relation of our notations and conventions to those of the functions
$U_n$ defined in \cite{Freitas} is given by:
\beq
(Q^2)^{-3\epsilon} \> {\bf F}(u,v,y,z) &=& -U_4(u,v,y,z),
\\
(Q^2)^{-3\epsilon} \> {\bf G}(w,u,z,v,y) &=& -U_5(u,z,v,y,w),
\\
(Q^2)^{-3\epsilon} \> {\bf H}(u,v,w,x,y,z) &=& U_6(u,v,y,z,x,w),
\eeq
where the \MSbar renormalization scale $Q$ is defined by eq.~(\ref{eq:defQ2}).

It is again useful to define $\epsilon$-independent modified basis 
integrals that will appear in renormalized quantities written 
in their most succinct forms. This is done by
subtracting UV sub-divergences and then taking the 4-dimensional limit. 
For the $\boldE$ and $\boldF$ integrals, 
we define:
\beq
E(u,v,y,z) &=& \lim_{\epsilon \rightarrow 0} \left[
\boldE(u,z,y,v)  
- \intE^{(1)}_{\rm div}(u,v,y,z)
- \intE^{(2)}_{\rm div}(u,v,y,z) 
- \intE^{(3)}_{\rm div}(u,v,y,z) 
\right ],
\eeq
where the 1-loop, 2-loop, and 3-loop UV sub-divergences are, respectively,
\beq
\intE^{(1)}_{\rm div}(u,v,y,z) &=& \frac{1}{\epsilon} 
\boldA(u) \boldA(v) + \mbox{(5 permutations)} ,
\\
\intE^{(2)}_{\rm div}(u,v,y,z) &=& 
\left [\frac{1}{2\epsilon^2}(v+y+z) +
\frac{1}{2\epsilon}\Bigl (\frac{u}{2}-v-y-z\Bigr ) \right ] \boldA(u) 
+ \mbox{(3 permutations)} ,
\\
\intE^{(3)}_{\rm div}(u,v,y,z) &=& 
\left [\frac{1}{3\epsilon^3} - \frac{2}{3\epsilon^2} + \frac{1}{3\epsilon} \right ]
(u v + u y + u z + v y + v z + y z)
\nonumber \\ &&
+
\left [\frac{1}{6 \epsilon^2} - \frac{3}{8\epsilon}  \right ]
(u^2 + v^2 + y^2 + z^2) .
\eeq
Then, renormalized
quantities can be written in terms of the function
\beq
F(u,v,y,z) = -\frac{\partial}{\partial u} E(u,v,y,z).
\label{eq:FfromE}
\eeq
From eq.~(\ref{eq:EEredundant}) and the other definitions above, one finds
the linear algebraic expression of the redundancy of $\intE$:
\beq
\intE (u,v,y,z) &=& \frac{1}{2} \Bigl [
-u \intF(u,v,y,z) - v \intF(v,u,y,z) - y \intF(y,u,v,z) - z \intF(z,u,v,y)
\nonumber \\ &&
+ \intA(u) \intA(v)
+ \intA(u) \intA(y)
+ \intA(u) \intA(z)
+ \intA(v) \intA(y)
+ \intA(v) \intA(z)
+ \intA(y) \intA(z)
\nonumber \\ &&
+ \left (u/2 -v-y-z\right ) \intA(u)
+ \left (v/2 -u-y-z\right ) \intA(v)
\nonumber \\ &&
+ \left (y/2 -u-v-z\right ) \intA(y)
+ \left (z/2 -u-v-y\right ) \intA(z)
\nonumber \\ &&
+ u v + u y + u z + v y + v z + y z
-9 (u^2 + v^2 + y^2 + z^2)/8 \Bigr ].
\label{eq:Eredundant}
\eeq
However, the function $\intF(u,v,y,z)$ has a logarithmic infrared divergence 
in the limit $u \rightarrow 0$. Therefore, we further define:
\beq
\Fbar(u,v,y,z) &\equiv& F(u,v,y,z) + \lnbar(u) \intI(v,y,z),
\label{eq:defFbar}
\eeq
which is well-defined for all finite values of its squared mass 
arguments. Some of the results described below are given in terms of the 
modified basis function $\Fbar$, and the program library {\tt 3VIL} uses 
$\Fbar$ rather than $\intF$ internally, but both functions 
are available as outputs, and eq.~(\ref{eq:defFbar}) can of course be 
used to translate between the $\intF$ and $\Fbar$ functions whenever 
necessary. In expressions below, 
we will use whichever of $\intF$ or $\Fbar$ is more convenient.

Similarly, we define the modified basis function:
\beq
\intG(w,u,z,v,y) &=& \lim_{\epsilon \rightarrow 0} \biggl[
\boldG(w,u,z,v,y)  
- \intG^{(1)}_{\rm div}(w,u,z,v,y)
- \intG^{(2)}_{\rm div}(w,u,z,v,y)
\nonumber \\ && 
- \intG^{(3)}_{\rm div}(w,u,z,v,y) 
\biggr ] ,
\label{eq:defintG}
\eeq
where the 1-loop, 2-loop, and 3-loop UV sub-divergences are 
\beq
\intG^{(1)}_{\rm div}(w,u,z,v,y) &=& \frac{1}{\epsilon} \left [
\boldI(w,u,z) + \boldI(w,v,y) \right ] ,
\\
\intG^{(2)}_{\rm div}(w,u,z,v,y) &=& 
\left (-\frac{1}{2\epsilon^2} + \frac{1}{2\epsilon} \right ) 
\left [\boldA(u) +  \boldA(v) + \boldA(y) + \boldA(z) \right ]
- \frac{1}{\epsilon^2} \boldA(w) ,
\\
\intG^{(3)}_{\rm div}(w,u,z,v,y) &=& 
\left (-\frac{1}{6\epsilon^3} + \frac{1}{2\epsilon^2} - \frac{2}{3\epsilon} \right )
(u+v+y+z)
%\nonumber \\ &&
+
\left (-\frac{1}{3\epsilon^3} + \frac{1}{3\epsilon^2} + \frac{1}{3\epsilon} \right )
w .
\eeq
A useful aspect of the definition eq.~(\ref{eq:defintG}) 
is that when renormalized expressions
are written in terms of $\intG$ rather than $\boldG$, then one does not need to use 
the $\epsilon^1$ parts of the expansions of $\boldI$ functions; 
only $I$ functions are necessary.

Finally, the $\boldH$ function is free of 1-loop and 2-loop UV sub-divergences, but does
have a 3-loop UV sub-divergence. Therefore we define:
\beq
H(u,v,w,x,y,z) &=& \lim_{\epsilon \rightarrow 0} \left[
\boldH(u,v,w,x,y,z)  
- \intH^{(3)}_{\rm div}(u,v,w,x,y,z)
\right ]
\eeq
where
\beq
\intH^{(3)}_{\rm div}(u,v,w,x,y,z) &=& 2 \zetathree/\epsilon .
\eeq
The function $H(u,v,w,x,y,z)$ 
is finite [except in the case $u=v=w=x=y=z=0$ where it has an infrared logarithmic divergence; see any one of eqs.~(\ref{eq:H00000x})-(\ref{eq:Hxxxxxx}) below].

By use of the integration by parts identities, 
the evaluation of a general 3-loop Feynman vacuum diagram is thus reduced to the problem
of computing $\intI(x,y,z)$, $\Fbar(u,v,y,z)$, $G(w,u,z,v,y)$, and $H(u,v,w,x,y,z)$. 
Although renormalized quantities are most 
efficiently written in terms of these quantities
rather than their bold-faced counterparts $\boldI(x,y,z)$,
$\boldF(u,v,y,z)$, $\boldG(w,u,z,v,y)$, and $\boldH(u,v,w,x,y,z)$, 
the formulas for the $\epsilon$-expansions
of the latter are provided in the next section.

The 2-loop integral $\intI(x,y,z)$ is known in terms of dilogarithms, 
but in general $\Fbar(u,v,y,z)$, $G(w,u,z,v,y)$, and $H(u,v,w,x,y,z)$ cannot be done analytically in terms of polylogarithms or other simple functions. 
Therefore, numerical methods are necessary.

\section{Expansions in $\epsilon$ for the integrals $\boldE$, $\boldF$, $\boldG$, 
and $\boldH$\label{sec:epsexp}}
\setcounter{equation}{0}
\setcounter{footnote}{1}

The $\epsilon$ expansions
of the $\boldE$, $\boldF$, $\boldG$, and $\boldH$ integrals
can be written in the forms:
\beq
\boldE(u,v,y,z) &=& 
\frac{1}{\epsilon^3} \Ethree(u,v,y,z)
+ \frac{1}{\epsilon^2}\Etwo(u,v,y,z)
+ \frac{1}{\epsilon}\Eone(u,v,y,z)
\nonumber \\ &&
+ \Ezero(u,v,y,z) + \ldots
\\
\boldF(u,v,y,z) &=& 
\frac{1}{\epsilon^3} \Fthree(u,v,y,z)
+ \frac{1}{\epsilon^2}\Ftwo(u,v,y,z)
+ \frac{1}{\epsilon}\Fone(u,v,y,z)
\nonumber \\ &&
+ \Fzero(u,v,y,z) + \ldots,
\\
\boldG(w,u,z,v,y) &=& 
\frac{1}{\epsilon^3} \Gthree(w,u,z,v,y)
+ \frac{1}{\epsilon^2}\Gtwo(w,u,z,v,y)
+ \frac{1}{\epsilon}\Gone(w,u,z,v,y)
\nonumber \\ &&
+ \Gzero(w,u,z,v,y) + \ldots\phantom{xxx}
\\
\boldH(u,v,w,x,y,z) &=& 
\frac{1}{\epsilon}\Hone(u,v,w,x,y,z)
+  \Hzero(u,v,w,x,y,z) + \ldots\phantom{xxx} .
\eeq
Using the formulas in section \ref{sec:basis}, one obtains:
\beq
\Ethree(u,v,y,z) &=& (u v + u y + u z + v y + v z + y z)/3,
\\
\Etwo(u,v,y,z) &=& 
- [
(v+y+z) \intA(u)
+ (u+y+z) \intA(v)
+ (u+v+z) \intA(y)
+ (u+v+y) \intA(z) ]/2
\nonumber \\ &&
+ (u v + u y + u z + v y + v z + y z)/3
- (u^2 + v^2 + y^2 + z^2)/12,
\\
\Eone(u,v,y,z) &=& 
\intA(u) \intA(v)
+ \intA(u) \intA(y)
+ \intA(u) \intA(z)
+ \intA(v) \intA(y)
+ \intA(v) \intA(z)
+ \intA(y) \intA(z)
\nonumber \\ &&
- (v+y+z) [\Aeps(u) + A(u)]/2
- (u+y+z) [\Aeps(v) + A(v)]/2
\nonumber \\ &&
- (u+v+z) [\Aeps(y) + A(y)]/2
- (u+v+y) [\Aeps(z) + A(z)]/2
\nonumber \\ &&
+ \bigl [u \intA(u) + v \intA(v) + y \intA(y) + z \intA(z) \bigr ]/4
\nonumber \\ &&
+ (u v + u y + u z + v y + v z + y z)/3
- 3 (u^2 + v^2 + y^2 + z^2)/8,
\\
\Ezero(u,v,y,z) &=& E(u,v,y,z)
+ \intA(u) \bigl[\Aeps(v) + \Aeps(y) + \Aeps(z)\bigr]
+ \intA(v) \bigl[\Aeps(u) + \Aeps(y) + \Aeps(z)\bigr]
\nonumber \\ &&
+ \intA(y) \bigl[\Aeps(u) + \Aeps(v) + \Aeps(z)\bigr]
+ \intA(z) \bigl[\Aeps(u) + \Aeps(v) + \Aeps(y)\bigr]
\nonumber \\ &&
- (v+y+z) \bigl[\Aeps(u) + \Aepstwo(u)\bigr]/2
- (u+y+z) \bigl[\Aeps(v) + \Aepstwo(v)\bigr]/2
\nonumber \\ &&
- (u+v+z) \bigl[\Aeps(y) + \Aepstwo(y)\bigr]/2
- (u+v+y) \bigl[\Aeps(z) + \Aepstwo(z)\bigr]/2
\nonumber \\ &&
+  [u \Aeps(u) + v \Aeps(v) + y \Aeps(y) + z \Aeps(z)]/4.
\eeq
Then one can use
\beq
F_n(u,v,y,z) &=& -\frac{\partial}{\partial u} E_n(u,v,y,z)
\eeq
for $n=0,1,2,3$, which can be evaluated using
\beq
\frac{\partial }{\partial u} \intA(u) &=& \intA(u)/u + 1,
\\
\frac{\partial }{\partial u} \Aeps(u) &=& [\Aeps(u) - \intA(u)]/u,
\\
\frac{\partial }{\partial u} \Aepstwo(u) &=& [\Aepstwo(u) - \Aeps(u)]/u,
\eeq
with the results:
\beq
\Fthree(u,v,y,z) &=& -(v+y+z)/3,
\\
\Ftwo(u,v,y,z) &=& 
(v + y + z) A(u)/2u
+ [A(v) + A(y) + A(z)]/2 +
(u + v + y + z)/6, \phantom{xxxxxxxxx}
\\
\Fone(u,v,y,z) &=& -[A(v) + A(y) + A(z)]A(u)/u + 
(v + y + z) \Aeps(u)/2u
\nonumber \\ &&
\!\!\!\!\!\!\!\!\!\!\!\!\!\!\!\!\!\!\!\!\!\!\!\!\!\!\!\!\!\!\!\!
+ [\Aeps(v) + \Aeps(y) + \Aeps(z) - A(u) - A(v) - A(y) - A(z)]/2
+ u/2 + (v+y+z)/6,
\\
\Fzero(u,v,y,z) &=& \intF(u,v,y,z)
+ (v + y + z) \Aepstwo(u)/2u 
- [A(v) + A(y) + A(z)] \Aeps(u)/u 
\nonumber \\ &&
\!\!\!\!\!\!\!\!\!\!\!\!\!\!\!\!\!\!\!\!\!\!\!\!\!\!\!\!\!\!\!\!
+ [A(v) + A(y) + A(z) - \Aeps(v) - \Aeps(y) - \Aeps(z) + u/4 
-v/2 - y/2 -z/2] A(u)/u
\nonumber \\ &&
\!\!\!\!\!\!\!\!\!\!\!\!\!\!\!\!\!\!\!\!\!\!\!\!\!\!\!\!\!\!\!\!
+ [\Aepstwo(v) + \Aepstwo(y) + \Aepstwo(z) - \Aeps(u) - \Aeps(v) - \Aeps(y) - \Aeps(z)]/2.
\eeq
Similarly, we obtain:
\beq
\Gthree(w,u,z,v,y) &=& -(2w+u+v+y+z)/6 ,
\\
\Gtwo(w,u,z,v,y) &=&  \bigl [\intA(u) + \intA(v) + \intA(y) + \intA(z) 
-u-v-y-z \bigr ]/2
+ \intA(w) - 2w/3 ,
\\
\Gone(w,u,z,v,y) &=& \intI(u,w,z) + \intI(v,w,y)  + \Aeps(w)
+ \bigl [\Aeps(u) + \Aeps(v) + \Aeps(y) + \Aeps(z) 
\nonumber \\ && 
+ \intA(u) + \intA(v) + \intA(y) + \intA(z) \bigr ]/2 + (w-2u-2v-2y-2z)/3,
\\
\Gzero(w,u,z,v,y) &=& \intG(w,u,z,v,y) + \Ieps(u,w,z) + \Ieps(v,w,y) - \Aepstwo(w)
+ \bigl [\Aeps(u) + \Aeps(v) 
\nonumber \\ && 
+ \Aeps(y) + \Aeps(z) 
-\Aepstwo(u) - \Aepstwo(v) - \Aepstwo(y) - \Aepstwo(z) \bigr ]/2 .
\eeq
Finally,
\beq
\Hone (u,v,w,x,y,z) &=& 2 \zeta(3),
\\
\Hzero(u,v,w,x,y,z) &=& H(u,v,w,x,y,z).
\eeq

Note that the 
the $\epsilon$-independent 
terms in the expansions, $\Ezero(u,v,y,z)$ and $\Fzero(u,v,y,z)$ and $\Gzero(w,u,z,v,y)$,
are not the same things as the more useful
functions $\intE(u,v,y,z)$ and $\intF(u,v,y,z)$ and $\intG(w,u,z,v,y)$. The latter appear
in renormalized quantities when put into the simplest forms.

%%%%%%%%%%%%%%%%%%%%%%%%%%%%%%%%%%%%%%%%%%%%%%%%%%%%%%%%%%%%%%%
\section{Derivatives of the basis functions \label{sec:massderivs}}
\setcounter{equation}{0}
\setcounter{figure}{0}
\setcounter{table}{0}
\setcounter{footnote}{1}

In this section, we give the derivatives of the basis functions
defined in the section \ref{sec:basis} with respect to the squared mass arguments
and the renormalization scale $Q$. These can be obtained using the 
integration by parts identities,
and are special cases of the general fact that any vacuum integral 
can be reduced to the basis. 
Note that the derivatives of ${\bf E}$ and $E$ functions are trivial, in 
the sense that they are just given by ${\bf F}$ and $F$ functions, 
respectively.

We start with the results in terms of the
bold-faced integrals ${\bf A}$, ${\bf I}$, ${\bf F}$, ${\bf G}$, and ${\bf H}$. 
For the 1-loop and 2-loop order basis integrals,
\beq
\frac{\partial}{\partial x} {\bf A}(x) &=& (d/2-1) {\bf A}(x)/x,
\\
\frac{\partial}{\partial x} {\bf I}(x,y,z) &=&
\Bigl \{
(d-3) (x-y-z) {\bf I}(x,y,z) + (d-2) \bigl [
(x-y+z) {\bf A}(x) {\bf A}(y)/2x
\nonumber \\ && 
+ (x+y-z) {\bf A}(x) {\bf A}(z)/2x
- {\bf A}(y) {\bf A}(z) \bigr ]
\Bigr \}/\lambda(x,y,z) ,
\eeq
where
\beq
\lambda (x,y,z) &\equiv& x^2 + y^2 + z^2 - 2 x y - 2 x z - 2 y z.
\label{eq:deflambda}
\eeq
The derivatives of ${\bf I}(x,y,z)$ with respect to $y$ and $z$ follow from symmetry.

For the 3-loop basis integrals, the results are more complicated, so that only the
structural forms will be shown in print here, 
with the complete explicit expressions relegated 
to an ancillary electronic file called {\tt derivatives.txt}, which is 
included with the arXiv source of this paper.
In all cases, the derivatives can be written as:
\beq
\sum_i k_i X_i
\eeq
where $X_i$ are basis integrals, and $k_i$ are rational functions of the squared masses
and the spacetime dimension $d$.
In the cases of 
\beq
\frac{\partial}{\partial u} {\bf F}(u,v,y,z) 
\eeq
\mbox{and}
\beq
\frac{\partial}{\partial v} {\bf F}(u,v,y,z) ,
\eeq
the basis integrals appearing in the sum are:
\beq
X_i &=&   \bigl \{ 
{\bf F}(u,v,y,z),\> 
{\bf F}(v,u,y,z),\> 
{\bf F}(y,u,v,z),\> 
{\bf F}(z,u,v,y) ,\> 
{\bf A}(u){\bf A}(v){\bf A}(y),\> 
\nonumber \\ && 
{\bf A}(u){\bf A}(v){\bf A}(z),\> 
{\bf A}(u){\bf A}(y){\bf A}(z),\> 
{\bf A}(v){\bf A}(y){\bf A}(z)
\bigr \}.
\eeq
The derivatives $\frac{\partial}{\partial y}{\bf F}(u,v,y,z)$ and
$\frac{\partial}{\partial z}{\bf F}(u,v,y,z)$ follow 
from $\frac{\partial}{\partial v}{\bf F}(u,v,y,z)$ by symmetry.
The denominators of the coefficients $k_i$ in these derivatives
contain factors of 
\beq
\psi(u,v,y,z) &\equiv&
u^4 + v^4 + y^4 + z^4 
- 4 u^3 (v+y+z)
- 4 v^3 (u+y+z)
- 4 y^3 (u+v+z)
\nonumber \\ &&
- 4 z^3 (u+v+y)
+ 4 u^2 (v y + v z + y z)
+ 4 v^2 (u y + u z + y z)
\nonumber \\ &&
+ 4 y^2 (u v + u z + v z)      
+ 4 z^2 (u y + u v + v y)  
+ 6 u^2 v^2 + 6 u^2 y^2 + 6 u^2 z^2 
\nonumber \\ &&
+ 6 v^2 y^2 
+ 6 v^2 z^2 + 6 y^2 z^2
- 40 u v y z 
.
\label{eq:defpsi}
\eeq
    
In the cases of
\beq
\frac{\partial}{\partial w} {\bf G}(w,u,z,v,y) 
\eeq
\mbox{and}
\beq
\frac{\partial}{\partial u} {\bf G}(w,u,z,v,y), 
\eeq
the basis integrals in the sum are:
\beq
X_i &=& \bigl \{ 
{\bf G}(w,u,z,v,y),\> 
{\bf F}(u,v,y,z),\>
{\bf F}(v,u,y,z),\>
{\bf F}(y,u,v,z),\>
{\bf F}(z,u,v,y),\>
\nonumber \\ &&
{\bf A}(v) {\bf I}(w,u,z),\>
{\bf A}(y) {\bf I}(w,u,z),\>
{\bf A}(u) {\bf I}(w,v,y),\>
{\bf A}(z) {\bf I}(w,v,y)
\bigr \}.
\eeq
The denominators of the coefficients for $\frac{\partial}{\partial w} {\bf G}(w,u,z,v,y)$
contain factors of $\lambda(u,w,z)$ and $\lambda(v,w,y)$, while the denominators
in $\frac{\partial}{\partial u} {\bf G}(w,u,z,v,y)$ contain only 
the factor $\lambda(u,w,z)$. The derivatives 
$\frac{\partial}{\partial z} {\bf G}(w,u,z,v,y)$,
$\frac{\partial}{\partial v} {\bf G}(w,u,z,v,y)$,
and
$\frac{\partial}{\partial y} {\bf G}(w,u,z,v,y)$
follow from 
$\frac{\partial}{\partial u} {\bf G}(w,u,z,v,y)$
using symmetry.

Finally, in the case of
\beq
\frac{\partial}{\partial u} {\bf H}(u,v,w,x,y,z) ,
\eeq
the necessary basis integrals are:
\beq
X_i &=& \bigl \{ 
{\bf H}(u,v,w,x,y,z),\> 
{\bf G}(u, v, x, w, z),\>
{\bf G}(v, u, x, w, y),\>
{\bf G}(w, u, z, v, y),\>
\nonumber \\ &&
{\bf G}(x, u, v, y, z),\>
{\bf G}(y, v, w, x, z),\>
{\bf G}(z, u, w, x, y),\>
{\bf F}(u, v, y, z),\>
{\bf F}(u, w, x, y),\>
\nonumber \\ &&
{\bf F}(v, u, y, z),\>
{\bf F}(v, w, x, z),\>
{\bf F}(w, u, x, y),\>
{\bf F}(w, v, x, z),\>
{\bf F}(x, u, w, y),\>
\nonumber \\ &&
{\bf F}(x, v, w, z),\>
{\bf F}(y, u, v, z),\>
{\bf F}(y, u, w, x),\>
{\bf F}(z, u, v, y),\>
{\bf F}(z, v, w, x),\>
\nonumber \\ &&
{\bf A}(w) {\bf I}(u, v, x),\>
{\bf A}(y) {\bf I}(u, v, x),\>
{\bf A}(z) {\bf I}(u, v, x),\>
{\bf A}(v) {\bf I}(u, w, z),\>
\nonumber \\ &&
{\bf A}(x) {\bf I}(u, w, z),\>
{\bf A}(y) {\bf I}(u, w, z),\>
{\bf A}(u) {\bf I}(v, w, y),\>
{\bf A}(x) {\bf I}(v, w, y),\>
\nonumber \\ &&
{\bf A}(z) {\bf I}(v, w, y),\>
{\bf A}(u) {\bf I}(x, y, z),\>
{\bf A}(v) {\bf I}(x, y, z),\>
{\bf A}(w) {\bf I}(x, y, z)\} .
\eeq
The denominators of the coefficients for 
$\frac{\partial}{\partial u} {\bf H}(u,v,w,x,y,z)$ contain factors of
$\lambda(u,v,x)$ and $\lambda(u,w,z)$ and
\beq
\chi(u,v,w,x,y,z)
&=&
u^2 y + v^2 z + w^2 x  + x^2 w + y^2 u + z^2 v
+ u v x - u w x - v w x 
\nonumber \\ && 
- u v y 
- u w y + v w y 
- u x y - w x y - u v z + u w z - v w z 
\nonumber \\ && 
- v x z - w x z - u y z - v y z + x y z
.
\label{eq:defchi}
\eeq
The derivatives of ${\bf H}(u,v,w,x,y,z)$ with respect to the other arguments follow 
from the tetrahedral symmetry.

The corresponding derivatives of the functions $A$, $I$, $F$, 
$\overline F$, $G$, and $H$ can be obtained straightforwardly 
from the results above and the formulas in the previous sections, 
by expanding in $\epsilon$. The results are quite complicated, 
so again they are not presented in print here, but are given explicitly 
in the ancillary file {\tt derivatives.txt}.  Where the denominator 
factors mentioned above vanish, the differential equations governing the 
basis functions have pseudo-thresholds, but the basis functions 
themselves are well-defined and smooth for all non-negative $u,v,w,x,y,z$.
 
It is also useful to have derivatives with respect to the 
renormalization scale, for example to check the renormalization group 
invariance of a calculation of the 3-loop effective potential. Here we 
present results in terms of the renormalized integrals $A$, $I$, $E$, 
$F$, $\Fbar$, $G$, and $H$. For the 1-loop and 2-loop integrals, one 
finds:
\beq
Q^2 \frac{\partial}{\partial Q^2} A (x) &=& -x,
\\
Q^2 \frac{\partial}{\partial Q^2} \intI (x,y,z) &=& A(x) + A(y) + A(z) -x-y-z.
\eeq
For the 4-propagator 3-loop integrals, we find:
\beq
Q^2 \frac{\partial}{\partial Q^2} \intE (u,v,y,z) &=&
A(u) A(v) + A(u) A(y) + A(u) A(z) + A(v) A(y) + A(v) A(z) + A(y) A(z)
\nonumber \\ && 
+ (u/2-v-y-z) A(u)
+ (v/2-u-y-z) A(v)
\nonumber \\ && 
+ (y/2-u-v-z) A(y)
+ (z/2-u-v-y) A(z)
\nonumber \\ && 
+ u v + u y + u z + v y + v z + y z
- 9 (u^2 + v^2 + y^2 + z^2)/8 ,
\\
Q^2 \frac{\partial}{\partial Q^2} \intF (u,v,y,z) &=&
[v +y +z -u -A(v) -A(y) - A(z)] A(u)/u + 7 u/4 ,
\\
Q^2 \frac{\partial}{\partial Q^2} \Fbar (u,v,y,z) &=&
A(v) + A(y) + A(z) - A(u) -\intI (v,y,z) - v-y-z + 7u/4   .
\eeq
For the 5- and 6-propagator 3-loop integrals, we obtain:
\beq
Q^2 \frac{\partial}{\partial Q^2} \intG (w,u,z,v,y) &=&
\intI (w,u,z) + \intI (w,v,y) + A(u) + A(v) + A(y) + A(z) 
\nonumber \\ && 
- 2 u - 2 v - 2 y - 2 z + w ,
\\
Q^2 \frac{\partial}{\partial Q^2} \intH (u,v,w,x,y,z) &=&
6 \zeta_3 .
\eeq

%%%%%%%%%%%%%%%%%%%%%%%%%%%%%%%%%%%%%%%%%%%%%%%%%%%%%%%%%%%%%%%
\section{Known analytical cases \label{sec:analytic}}
\setcounter{equation}{0}
\setcounter{figure}{0}
\setcounter{table}{0}
\setcounter{footnote}{1}

For some special cases, it is possible to give analytical expressions in closed
form for the basis integrals, in terms of the polylogarithm functions ${\rm Li}_n(z)$
of complex argument \cite{Lewin}. Although individual terms in expressions below are 
sometimes complex numbers, the basis vacuum integrals are always real when the 
squared masses are non-negative. Besides the usual
transcendental numbers such as $\ln(2)$, $\pi$, 
$\zeta_3$ and ${\rm Li}_4(1/2)$, some expressions below
involve the log-sine definite integrals:
\beq
\Lstwo &\equiv& \Lstwo(2\pi/3) = -\int_0^{2\pi/3} dx \ln [2 \sin(x/2)] \>\approx\> 
 0.6766277376064358 ,
\\
\Lsthree &\equiv& \Lsthree(2\pi/3) =  -\int_0^{2\pi/3} dx \ln^2 [2 \sin(x/2)] \>\approx\> 
-2.1447672125694944 ,
\\
\Lspfour &\equiv& {\rm Ls}_4^{(1)}(2\pi/3)  = -\int_0^{2\pi/3} dx \,x \ln^2 [2 \sin(x/2)] \>\approx\> 
-0.4976755516066472 .
\eeq
The function $\Lstwo(x)$ is also known as the Clausen function of
order 2, and is often denoted instead as ${\rm Cl}_2(x)$.

The 2-loop vacuum integral basis function $\intI(x,y,z)$ is well-known, in various
cosmetically different but equivalent forms
\cite{Ford:1992pn,Davydychev:1992mt,Davydychev:1995mq,Caffo:1998du,Espinosa:2000df}.
For $z \geq x,y$:
\beq
\intI(x,y,z) &=& s \Bigl [\dilog(k_1) + \dilog(k_2)
- \ln(k_1)  \ln(k_2) + \frac{1}{2} \ln(x/z) \ln(y/z) - \pi^2/6 \Bigr ]
\nonumber \\ &&
+ \frac{1}{2} (z-x-y) \lnbar(x) \lnbar(y)
+ \frac{1}{2} (y-x-z) \lnbar(x) \lnbar(z)
+ \frac{1}{2} (x-y-z) \lnbar(y) \lnbar(z)
\nonumber \\ &&
+ 2 x \lnbar(x) + 2 y \lnbar(y) + 2z \lnbar(z)
- \frac{5}{2}(x+y+z)
\label{eq:analI}
\eeq
where 
\beq
s &=&  \sqrt{\lambda(x,y,z)},
\label{eq:definesI2}
\\
k_1 &=& (x+z-y-s)/2z,
\\
k_2 &=& (y+z-x-s)/2z.
\eeq
The cases with $y\geq x,z$ or $x \geq y,z$ are obtained by permuting the 
arguments of eq.~(\ref{eq:analI}). Some useful special cases are:
\beq
\intI(0,y,z) &=& (y-z) \Bigl [ \dilog(1-y/z) + \frac{1}{2} \lnbar^2(z) \Bigr ]
- y \lnbar(y) \lnbar(z) 
\nonumber \\ && 
+ 2 y \lnbar(y) + 2 z \lnbar(z) - \frac{5}{2}(y+z),\phantom{xxx}
\\
\intI (x,x,x) &=& x \left [-\frac{15}{2} + 3 \sqrt{3} \Lstwo + 6 \lnbar(x) 
- \frac{3}{2} \lnbar^2(x) \right ] 
\label{eq:Ixxx}
,
\\
\intI(0,x,x) &=& x \left [-5 + 4 \lnbar(x) - \lnbar^2(x) \right ]
,
\\
\intI(0,0,x) &=& x \left [-\frac{5}{2} - \frac{\pi^2}{6} 
+ 2 \lnbar(x) - \frac{1}{2} \lnbar^2(x) \right ]
,
\\
\intI(0,0,0) &=& 0.
\eeq
Now the results for $\Izero(x,y,z)$ can be obtained easily from eqs.~(\ref{eq:defAeps})
and (\ref{eq:relateII0}). 

The result for $\Ieps(x,y,z)$ can be obtained as 
a straightforward application of the method in ref.~\cite{Ford:1992pn}, and 
has been given in a more compact form in eqs.~(15)-(21) and (41) 
of the preprint version of ref.~\cite{Davydychev:1995mq}, based on functions
defined in eqs.~(11), (12), and (29) of ref.~\cite{Usyukina:1994iw}.
(See also ref.~\cite{Davydychev:1999mq} 
for the expansion of ${\bf I}(x,y,z)$ to all orders in
$\epsilon$.) 
These results for $\Ieps(x,y,z)$ take different forms depending on 
whether $x+y$ is greater or less than $z$. However,
the results can be rewritten in a unified way for all $z\geq x,y$ with $s \not=0$ and $z\not=x+y$, as:
\beq
\Ieps(x,y,z) &=&
[3 - \lnbar(x) - \lnbar(y)] \intI(x,y,z) + 
[x \lnbar^3(x) + y \lnbar^3(y) + z \lnbar^3(z)]/6
\nonumber \\ &&
+ [x \lnbar^2(x) + y \lnbar^2(y) -3 z \lnbar^2(z)]/2
+ [(y-x-z)/4] \lnbar(x) \lnbar(z) \ln(x/z)
\nonumber \\ &&
+ [(x-y-z)/4] \lnbar(y) \lnbar(z) \ln(y/z)
+ [(z-x-y)/4] \lnbar(x) \lnbar(y) [\lnbar(x) + \lnbar(y)]
\nonumber \\ &&
+ [2x + 2 y - z \lnbar(z)] \lnbar(x) \lnbar(y)
+ 2 z \lnbar(z) [\lnbar(x) + \lnbar(y)]
+ (\pi^2/6 + 1) z \lnbar(z)
\nonumber \\ &&
+ (\pi^2/6 - 3/2) [x \lnbar(x) + y \lnbar(y)] 
-5 [(y+z) \lnbar(x) + (x+z) \lnbar(y)]/2
\nonumber \\ &&
+ (\zeta_3/3 - \pi^2/4) (x+y+z)
+ s \Bigl \{
\trilog(1-r_x) 
+ \trilog(1-r_y) 
+ \trilog(1-r_z)
\nonumber \\ &&
- \trilog(1-1/r_x)
- \trilog(1-1/r_y)
- \trilog(1-1/r_z)
+ \ln(z/x) \dilog(1-r_x)
\nonumber \\ &&
+ \ln(z/y) \dilog(1-r_y)
+ \frac{1}{4} \ln(r_x) \ln(r_y) \ln(r_z)
+ \frac{1}{4} \ln(z/x) \ln(r_x) \ln(x r_x/z)
\nonumber \\ &&
+ \frac{1}{4} \ln(z/y) \ln(r_y) \ln(y r_y/z)
+ \frac{\eta}{4} \Bigl [\ln^2(-s^2/xy) - \ln(r_x) \ln(r_y) 
\nonumber \\ &&+ [\ln(r_x) + \ln(r_y) - \ln(r_z)]^2/4 \Bigr ]
\Bigr \} ,
\label{eq:Iepsanal}
\eeq
where $s$ was defined above in eq.~(\ref{eq:definesI2}), and
\beq
r_x &=& (s+x-y-z)^2/4 y z,
\\
r_y &=& (s+y-x-z)^2/4 x z,
\\
r_z &=& (s+z-x-y)^2/4 x y,
\eeq
which implies that $r_x r_y r_z = 1$, and
\beq
\eta &\equiv& \ln(r_x) + \ln(r_y) + \ln(r_z) = 
\Biggl \{ \begin{array}{ll}
-2\pi i & \quad \mbox{(for $x+y<z$),}
\\
0 & \quad \mbox{(for $x+y>z$).}
\end{array}
\Biggr.
\label{eq:defeta}
\eeq
The special case with $s =0$ is obtained by simply removing
all of the terms multiplied by $s$ (i.e., the ones enclosed in curly brackets) 
in eq.~(\ref{eq:Iepsanal}). The special case $z = x+y$ can be computed 
by taking the limit 
$z \rightarrow x+y$ of eq.~(\ref{eq:Iepsanal}), either from above or from below;
these limits coincide, despite the branch cut 
discontinuity in eq.~(\ref{eq:defeta}).
Other mass orderings $x \geq y,z$ or $y \geq x,z$ 
are obtained by permuting the arguments
of eq.~(\ref{eq:Iepsanal}). 
Some useful special cases are:
\beq
\Ieps(0,x,y) &=&
(y-x) \Bigl \{
{\rm Li}_3 (1-x/y) - {\rm Li}_3 (1-y/x) +
[\lnbar(x) +\lnbar(y) - 3] {\rm Li}_2 (1-x/y)
\Bigr \} 
\nonumber \\ &&
+ (y/6) \lnbar^3(x) 
+ \left ( x - y/2 \right ) \lnbar^2(x) \lnbar(y) 
+ (y/2) \lnbar(x) \lnbar^2(y) 
\nonumber \\ &&
+ \left (y/2 - x/3 \right ) \lnbar^3(y)
- (3x/2) \lnbar^2(x)
- 3 x \lnbar(x) \lnbar(y) 
+ \left ( 3x/2 - 3 y \right ) \lnbar^2(y)
\nonumber \\ &&
+ \left (7 + \pi^2/6 \right ) [x \lnbar(x) + y \lnbar(y)]
+ \left (\zeta_3/3 - 15/2 - \pi^2/4 \right ) (x+y)
,
\\
\Ieps(x,x,x) &=& x \Bigl \{
2 \lnbar^3(x) - 9 \lnbar^2(x)
+ \bigl [21 + \pi^2/2 - 6 \sqrt{3} \Lstwo \bigr ]\lnbar(x)
- 45/2 - 3 \pi^2/4 
\nonumber \\ &&
+ [3  - \ln(3)] 3 \sqrt{3} \Lstwo 
+ 3 \sqrt{3} \Lsthree + \zeta_3 + \pi^3/2\sqrt{3}
\Bigr \}
,
\\
\Ieps(0,x,x) &=& x \Bigl [
\frac{4}{3} \lnbar^3(x) - 6 \lnbar^2(x) 
+ \Bigl ( 14 + \frac{\pi^2}{3} \Bigr )\lnbar(x)
- 15 - \frac{\pi^2}{2} + \frac{2 \zeta_3}{3}
\Bigr ]
,
\\
\Ieps(0,0,x) &=& x \Bigl [
\frac{2}{3} \lnbar^3(x) - 3 \lnbar^2(x) 
+ \Bigl ( 7 + \frac{\pi^2}{2} \Bigr )\lnbar(x)
- \frac{15}{2} - \frac{3\pi^2}{4} + \frac{4 \zeta_3}{3}
\Bigr ]
,
\\
\Ieps(0,0,0) &=& 0
.
\label{eq:Ieps000}
\eeq

The results for the 3-loop integrals $\boldE$, $\boldF$, $\boldG$, $\boldH$ 
involving propagators that are either
massless or contain a single non-zero mass scale
were obtained in \cite{Broadhurst:1991fi,
Avdeev:1994db,
Fleischer:1994dc,
Avdeev:1995eu,
Broadhurst:1998rz,
Fleischer:1999mp,
Schroder:2005va}.
A particularly useful and systematic source for them is found in 
\cite{Schroder:2005va}. For convenience, we provide below 
these results in terms of our modified functions $\intE,\intF,\Fbar,\intG,\intH$. 
The expansions of
$\boldE$, $\boldF$, $\boldG$, $\boldH$ up through order $\epsilon^0$ can 
be reconstructed from these results, using the results of 
section \ref{sec:epsexp} of the present paper.

The special cases involving four propagators with all propagator 
squared masses equal to either 0 or $x$ include:
\beq
\intE(0,0,0,0) &=& 0,
\\
\intE(0,0,0,x) &=& x^2 \left [-\frac{133}{48}-\frac{\pi^2}{12} + 
\frac{13}{8}\lnbar(x) - \frac{1}{4}\lnbar^2(x) \right ],
\\
\intE(0,0,x,x) &=& x^2 \left [
\frac{8 \zeta_3}{3} -\frac{89}{24} - \frac{3}{4} \lnbar(x) + 
\frac{3}{2} \lnbar^2(x)
- \frac{1}{3} \lnbar^3(x)
\right ],
\\
\intE(0,x,x,x) &=& x^2 \left [
\frac{9\sqrt{3}}{2} \Lstwo - \frac{45}{16} - \frac{57}{8} \lnbar(x) + 
\frac{21}{4} \lnbar^2(x) - \lnbar^3(x) \right ] ,
\\
\intE(x,x,x,x) &=& x^2 \left [
-\frac{1}{12} - \frac{35}{2} \lnbar(x) + 11 \lnbar^2(x) - 2 \lnbar^3(x)
\right ],
\eeq
and
\beq
\intF(x,0,0,0) &=& 
x \left [\frac{47}{12} + \frac{\pi^2}{6} 
- \frac{11}{4} \lnbar(x) 
+ \frac{1}{2} \lnbar^2(x) \right ] ,
\\
\intF(x,0,0,x) &=& 
x \left [\frac{49}{12} - \frac{8}{3}\zeta_3 
- \frac{3}{4} \lnbar(x) 
- \lnbar^2(x) + \frac{1}{3} \lnbar^3(x) \right ] ,
\\
\intF(x,0,x,x) &=& 
x \left [\frac{17}{4} - 3 \sqrt{3} \Lstwo 
+\frac{5}{4} \lnbar(x) 
- \frac{5}{2} \lnbar^2(x) + \frac{2}{3} \lnbar^3(x) \right ] ,
\\
\intF(x,x,x,x) &=&
x \left [\frac{53}{12} + \frac{13}{4} \lnbar(x) 
-4 \lnbar^2(x) + \lnbar^3(x) \right ] ,
\eeq
and
\beq
\Fbar(0,0,0,0) &=& 0,
\\
\Fbar(0,0,0,x) &=& 
x \left [\frac{1}{6} + \frac{\pi^2}{6} - \frac{2}{3} \zeta_3 
- \left (\frac{1}{2} + \frac{\pi^2}{6}\right ) \lnbar(x) 
+ \frac{1}{2} \lnbar^2(x) - \frac{1}{6} \lnbar^3(x) \right ] 
,
\\
\Fbar(0,0,x,x) &=& 
x \left [\frac{1}{3} + \frac{8}{3}\zeta_3 
- \lnbar(x) 
+ \lnbar^2(x) - \frac{1}{3} \lnbar^3(x) \right ] ,
\\
\Fbar(x,0,0,0) &=& 
x \left [\frac{47}{12} + \frac{\pi^2}{6} 
- \frac{11}{4} \lnbar(x) 
+ \frac{1}{2} \lnbar^2(x) \right ] ,
\\
\Fbar(x,0,0,x) &=& 
x \left [\frac{49}{12} - \frac{8}{3}\zeta_3 
- \left (\frac{13}{4} + \frac{\pi^2}{6} \right ) \lnbar(x) 
+ \lnbar^2(x) - \frac{1}{6} \lnbar^3(x) \right ] ,
\\
\Fbar(x,0,x,x) &=& 
x \left [\frac{17}{4} - 3 \sqrt{3} \Lstwo 
-\frac{15}{4} \lnbar(x) 
+ \frac{3}{2} \lnbar^2(x) - \frac{1}{3} \lnbar^3(x) \right ] ,
\\
\Fbar (x,x,x,x) &=& 
x \left [\frac{53}{12} + 
\left (3 \sqrt{3} \Lstwo - \frac{17}{4}\right ) \lnbar(x) 
+ 2 \lnbar^2(x) - \frac{1}{2} \lnbar^3(x) \right ] 
,
\label{eq:Fbarxxxx}
\eeq
and others obtained by permutations implied by the symmetries of the graphs.
There is only one such case for which we do not know an exact analytic expression:\footnote{Note added in v3, August 8, 2021:
after the publication of this paper, we have determined that the exact analytical 
form is
$9.09686753726327768\ldots = \frac{1}{2} + 3 \sqrt{3}(2 \ln 3 - 1)  \Lstwo - 6 \sqrt{3} \Lsthree - \pi^3/\sqrt{3}$.}
\beq
\Fbar(0,x,x,x) &\approx& x \left [9.09686753726327768\ldots
+ (3 \sqrt{3} \Lstwo - 3/2) \lnbar(x) 
+ \frac{3}{2} \lnbar^2(x) - \frac{1}{2} \lnbar^3(x) \right ] .\phantom{xxx}
\eeq
Here, the  numerical part was found using high-order series solutions of
the differential equation. The cases with five or six propagators that are all the same or 0 are:
\beq
\intG(0,0,0,0,0) &=& 0,
\\
\intG(0,0,0,0,x) &=& 
x \left [-\frac{15}{2} - \frac{\pi^2}{2} + \frac{2}{3} \zeta_3 + 
\left (\frac{11}{2} + \frac{\pi^2}{6} \right ) \lnbar(x) 
- \frac{3}{2} \lnbar^2(x) + \frac{1}{6} \lnbar^3(x) \right ] ,
\\
\intG(x,0,0,0,0) &=& 
x \left [-\frac{7}{3} - \frac{2\pi^2}{3} - \frac{2}{3} \zeta_3 + 
\left (4 + \frac{\pi^2}{3} \right ) \lnbar(x) 
- 2 \lnbar^2(x) + \frac{1}{3} \lnbar^3(x) \right ] ,
\\
\intG(0,0,0,x,x) &=& 
x \left [-15 - \frac{8}{3} \zeta_3 + 
11 \lnbar(x) 
- 3 \lnbar^2(x) + \frac{1}{3} \lnbar^3(x) \right ] ,
\\
\intG(0,0,x,0,x) &=& 
x \left [-15 - \frac{\pi^2}{3} + \frac{16}{3} \zeta_3 + 
\left (11 + \frac{\pi^2}{3} \right ) \lnbar(x) 
- 3 \lnbar^2(x) + \frac{1}{3} \lnbar^3(x) \right ] ,
\\
\intG(x,0,0,0,x) &=& 
x \left [-\frac{59}{6} - \frac{\pi^2}{2} + 
\left (\frac{19}{2} + \frac{\pi^2}{6} \right ) \lnbar(x) 
- \frac{7}{2} \lnbar^2(x) + \frac{1}{2} \lnbar^3(x) \right ] ,
\\
\intG(0,0,x,x,x) &=& 
x \left [-\frac{45}{2} + 9 \sqrt{3} \Lstwo + 
\left (\frac{33}{2} + \frac{\pi^2}{6} \right ) \lnbar(x) 
- \frac{9}{2} \lnbar^2(x) + \frac{1}{2} \lnbar^3(x) \right ] ,
\\
\intG(x,0,0,x,x) &=& 
x \Bigl [-\frac{52}{3} + 6 \sqrt{3} \Lstwo - \frac{\pi^2}{3} - 
\frac{2 \pi^3}{9\sqrt{3}} 
- \frac{4}{3}\zeta_3 + 
\left (15 + \frac{\pi^2}{6} - 3 \sqrt{3} \Lstwo \right ) \lnbar(x) 
\nonumber \\ &&
- 5 \lnbar^2(x) + \frac{2}{3} \lnbar^3(x) \Bigr ] ,
\\
\intG(x,0,x,0,x) &=& 
x \Bigl [-\frac{52}{3} + \frac{8}{3}\zeta_3 + 15 \lnbar(x) 
- 5 \lnbar^2(x) + \frac{2}{3} \lnbar^3(x) \Bigr ] ,
\\
\intG(0,x,x,x,x) &=& 
x \Bigl [-30 + \frac{56}{3}\zeta_3 + 22 \lnbar(x) 
- 6 \lnbar^2(x) + \frac{2}{3} \lnbar^3(x) \Bigr ] ,
\\
\intG(x,0,x,x,x) &=& 
x \Bigl [-\frac{149}{6} + 9 \sqrt{3} \Lstwo + 
\left (\frac{41}{2} - 3 \sqrt{3} \Lstwo \right ) \lnbar(x) 
- \frac{13}{2} \lnbar^2(x) + \frac{5}{6} \lnbar^3(x) \Bigr ] ,
\\
\intG (x,x,x,x,x) &=& 
x \left [-\frac{97}{3} + 12 \sqrt{3} \Lstwo + 6 \zetathree +
\left (26 - 6 \sqrt{3} \Lstwo \right ) \lnbar(x) 
-8 \lnbar^2(x) + \lnbar^3(x) \right ] 
,
\phantom{xxxx}
\label{eq:Gxxxxx}
\eeq
and
\beq
\intH (0,0,0,0,0,x) &=& 
\frac{\pi^4}{30} + 6 \zeta_3 [1 - \lnbar(x)],
\label{eq:H00000x}
\\
\intH (0,0,0,0,x,x) &=& 
-\frac{\pi^4}{18} + 6 \zeta_3 [1 - \lnbar(x)],
\\
\intH (0,x,0,0,0,x) &=&
16 {\rm Li}_4(1/2) - \frac{7\pi^4}{60}
+ \frac{2}{3} \ln^2(2) [\ln^2(2) - \pi^2]
+6 \zetathree [1 - \lnbar(x)] ,  
\\
\intH (0,0,0,x,x,x) &=&
-\frac{11\pi^4}{180} - 9 (\Lstwo)^2
+6 \zetathree [1 - \lnbar(x)] ,  
\\
\intH (0,0,x,0,x,x) &=&
-\frac{\pi^4}{10}
+6 \zetathree [1 - \lnbar(x)] ,  
\\
\intH (0,0,x,x,0,x) &=&
-\frac{\pi^4}{24} - \frac{27}{2} (\Lstwo)^2
+6 \zetathree [1 - \lnbar(x)] ,  
\\
\intH (0,0,x,x,x,x) &=&
-\frac{77\pi^4}{1080} - \frac{27}{2} (\Lstwo)^2
+6 \zetathree [1 - \lnbar(x)] ,  
\\
\intH (0,x,x,x,0,x) &=&
32 {\rm Li}_4(1/2) - \frac{11\pi^4}{45}
+ \frac{4}{3} \ln^2(2) [\ln^2(2) - \pi^2]
+6 \zetathree [1 - \lnbar(x)] ,  
\\
\intH (0,x,x,x,x,x) &=&
\frac{7\pi^4}{30} - 6 (\Lstwo)^2 + 4 \pi \Lsthree - 6 \Lspfour 
- \frac{26}{3}\ln(3)\zetathree
+6 \zetathree [1 - \lnbar(x)] ,
\\
\intH (x,x,x,x,x,x) &=&
16 {\rm Li}_4(1/2) - \frac{17\pi^4}{90}
+ \frac{2}{3} \ln^2(2) [\ln^2(2) - \pi^2]
- 9 (\Lstwo)^2
+6 \zetathree [1 - \lnbar(x)]   
,\phantom{xxxx}
\label{eq:Hxxxxxx}
\eeq
and others obtained by permutations implied by the symmetries of the graphs.

Some cases involving two distinct non-zero masses can also be given analytically.
Equation (4.19) of ref.~\cite{Davydychev:2003mv} gives $\boldH(0,0,x,y,x,x)$ through
order $\epsilon^0$,
in terms of Nielsen generalized polylogarithm functions, 
and eq.~(3.27) of ref.~\cite{Kalmykov:2005hb} provides the $\epsilon$ expansion of
$\boldG(y,x,x,x,x)$ in terms of log-sine integrals. For brevity,
those results are omitted here. 
Reference~\cite{Bytev:2009mn} obtained the equivalent of $\boldE(x,x,y,y)$ 
and $\boldF(x,x,y,y)$ to all orders in $\epsilon$
in terms of hypergeometric functions.
Reference~\cite{Bekavac:2009gz} obtained the equivalent of
$E(x,x,y,y)$ and $F(x,x,y,y)$.
Reference~\cite{Bytev:2009kb}
contains the expansions of $\boldG(x,0,0,0,y)$ 
and $\boldG(x,0,0,y,y)$ and $\boldE(0,x,x,y)$ and $\boldF(x,0,x,y)$,
in eqs.~(64), (81), (90), and (90) respectively, while
ref.~(\cite{Bytev:2011ks}) contains an expression for $\boldG(x,0,0,0,y)$ to all orders
in $\epsilon$ in terms of hypergeometric functions.
Reference~\cite{Kalmykov:2006pu} found results for $\boldE(0,0,x,y)$ and $\boldF(0,0,x,y)$
and $\boldF(x,0,0,y)$ to all orders in $\epsilon$ in terms of hypergeometric functions.
Reference~\cite{Grigo:2012ji} also found results for the $\epsilon$ expansions of
the equivalents of $\boldE(0,0,x,y)$ and $\boldF(x,0,0,y)$
in terms of harmonic polylogarithms.
Each of those results can be written in terms of only ordinary polylogarithms 
up through order $\epsilon^0$. 
We have also solved the differential equations to obtain a few more cases involving
two distinct non-zero masses. Below we list only the cases that can be written in terms of
ordinary polylogarithms. This includes the 4-propagator cases:
\beq
\intE (0,0,x,y) &=&
x y \bigl [-2 \trilog(1-x/y) -2 \trilog(1-y/x) 
+ (1/3) \lnbar^3(x) - (1/6) \lnbar^3(y) 
\nonumber \\ &&
- \lnbar^2(x) \lnbar(y)
+ (1/2) \lnbar(x) \lnbar^2(y) 
+ 2 \lnbar(x) \lnbar(y) 
- 2 \lnbar(x) 
- 2 \lnbar(y)
+ 8 \zetathree/3  
\nonumber \\ &&
+ 11/6 
\bigr ]
+[(x^2-y^2)/2 + x y \ln(x/y) ] \dilog(1-x/y) 
- (x^2/2) \lnbar(x) \lnbar(y)
\nonumber \\ &&
+ [(x^2 - y^2)/4] \lnbar^2(y)
+ (13 x^2/8) \lnbar(x)
+ (13 y^2/8) \lnbar(y)
-133 (x^2 + y^2)/48 ,
\label{eq:E00xy}
\\
\intF (x,0,0,y) &=&
2 y \trilog(1-x/y) + 2 y \trilog(1-y/x) + [y \ln(y/x) - x + y] \dilog(1-x/y)
 \nonumber \\ &&
-(y/3) \lnbar^3(x)
+(y/6) \lnbar^3(y)
+ y \lnbar^2(x) \lnbar(y)
-(y/2) \lnbar(x) \lnbar^2(y)
 \nonumber \\ &&
+ (x-2y) \lnbar(x) \lnbar(y)
+ [(y-x)/2] \lnbar^2(y)
+ (5y/2 - 11x/4) \lnbar(x)
 \nonumber \\ &&
- (y/2) \lnbar(y)
+ 47x/12 + y/6- 8 \zetathree y/3 
,
\label{eq:Fx00y}
\\
\Fbar (0,0,x,y) &=&
-2 x \trilog(1-y/x) - 2 y \trilog(1-x/y)
+ \left [x \lnbar(x) - y \lnbar(y) -x + y \right ] \dilog(1-x/y)
 \nonumber \\ &&
+ (x/3) \lnbar^3(x)
- (y/6) \lnbar^3(y)
- x \lnbar^2(x) \lnbar(y)
+ (x/2) \lnbar(x) \lnbar^2(y)
+ x \lnbar(x) \lnbar(y)
 \nonumber \\ &&
+ [(y-x)/2] \lnbar^2(y)
- (x/2) \lnbar(x)
- (y/2) \lnbar(y)
+ (x + y) (1/6 + 4 \zetathree/3) ,
\\
\intE (x,x,y,y) &=&
-2 (x-y)^2 \bigl[ \trilog(1-x/y) + \trilog(1-y/x) + \ln(y/x) \dilog(1-x/y)
\nonumber \\ &&
- (1/6) \lnbar^3(x) + (1/3) \lnbar^3(y) - 7\zetathree/3 \bigr] 
+ (2 x y - 2 x^2 - y^2) \lnbar^2(x) \lnbar(y)
\nonumber \\ &&
+ (2 x^2 - 4 x y + y^2) \lnbar(x) \lnbar^2(y)
+ x (3x/2 -y) \lnbar^2(x)
+ y (3y/2 -x) \lnbar^2(y)
\nonumber \\ &&
+ 10 x y \lnbar(x) \lnbar(y)
- x (8 y + 3x/4) \lnbar(x)
- y (8 x + 3y/4) \lnbar(y)
\nonumber \\ &&
-89(x^2 + y^2)/24 + 22 x y/3,
\label{eq:Exxyy}
\\
\intF (x,x,y,y) &=&
2 (x-y) \bigl [\trilog(1-x/y) + \trilog(1-y/x) + \ln(y/x) \dilog(1-x/y) 
-(1/6) \lnbar^3(x)
 \nonumber \\ &&
+ (1/3) \lnbar^3(y)
- \lnbar(x) \lnbar^2(y)
- 7 \zetathree/3 \bigr ]
+ (2x-y) \lnbar^2(x) \lnbar(y)
- x \lnbar^2(x)
 \nonumber \\ &&
- 4 y \lnbar(x) \lnbar(y)
+ y \lnbar^2(y)
+ (5 y - 3 x/4) \lnbar(x)
- y \lnbar(y)
+ 49 x/12 + y/3
,
\label{eq:Fxxyy}
\\
\intE (0,x,y,y) &=&
y(y-x) \Bigl [ 
4 \trilog(-k)  + (1/3) \ln^3(k) - (1/3) \lnbar^3(y) + 
(\pi^2/3) \ln(k) - 4 \zetathree/3
\Bigr ]
\nonumber \\ &&
- (x+2 y) \sqrt{x^2 - 4 x y} \Bigl [
\dilog(-k) + (1/4) \ln^2(k) + \pi^2/12
\Bigr ]
+ [(x^2 + 6 y^2)/4] \lnbar^2(y) 
\nonumber \\ &&
- x y \lnbar(x) \lnbar^2(y) 
+ [x (8y-x)/2] \lnbar(x) \lnbar(y)
+ [x (13 x - 32 y)/8] \lnbar(x)
\nonumber \\ &&
- [y (16 x + 3 y)/4] \lnbar(y) 
-133 x^2/48 + 11 x y/3 - 89 y^2/24,
\label{eq:E0xyy}
\\
\intF (x,0,y,y) &=&
4 y \trilog(-k) 
+ \sqrt{x^2 - 4 x y} \Bigl [ 2 \dilog(-k)
+ (1/2) \ln^2(k) + \pi^2/6 \Bigr ]
\nonumber \\ &&
+ (y/3) \bigl [
\ln^3(k) - \lnbar^3(y) + \pi^2 \ln(k) + 1 - 4 \zetathree + 
3 \lnbar(x) \lnbar^2(y) - 3 \lnbar(y) \bigr ]
\nonumber \\ &&
+ (y-x/2) \lnbar^2(y) + (x-4y) \lnbar(x) \lnbar(y) +
(5y - 11x/4) \lnbar(x) + 47 x/12,
\label{eq:Fx0yy}
\\
\intF (y,0,y,x) &=&
(x-2y) \Bigl [ 
2 \trilog(-k) + (1/6) \ln^3(k) - (1/6) \lnbar^3(y)
+ (\pi^2/6) \ln(k) -  2\zetathree/3
\Bigr ]
\nonumber \\ &&
+  \sqrt{x^2 - 4 x y} \bigl [ 2 \dilog(-k)
+ (1/2) \ln^2(k) + \pi^2/6 \bigr ]
+ (x/2) \lnbar(x) [\lnbar^2(y) -1]
\nonumber \\ &&
- x \lnbar(x) \lnbar(y) - (y + x/2) \lnbar^2(y) 
+ (5x/2 - 3y/4) \lnbar(y) + x/6 + 49 y/12 
,
\label{eq:Fy0yx}
\eeq
with, in the last three equations,
\beq
k &\equiv& \bigl (1 - \sqrt{1-4y/x} \bigr )/\bigl (1 + \sqrt{1-4y/x} \bigr ).
\label{eq:defk}
\eeq
Equations (\ref{eq:E00xy}) and (\ref{eq:Fx00y}) 
are equivalent to results already found by ref.~\cite{Grigo:2012ji}.
Equations (\ref{eq:Exxyy}) and (\ref{eq:Fxxyy}) are equivalent to results obtained 
by refs.~\cite{Bytev:2009mn} and \cite{Bekavac:2009gz}.
Equations (\ref{eq:E0xyy}), (\ref{eq:Fx0yy}) and (\ref{eq:Fy0yx}) 
are equivalent to results 
already found in eq.~(90) of ref.~\cite{Bytev:2009kb}.
Of course, the corresponding $\Fbar$ integrals
can also be obtained from the results above, using eq.~(\ref{eq:defFbar}).

The 5-propagator integrals with two distinct non-zero masses 
that we have been able to find analytically in terms of ordinary polylogarithms
are:
\beq
\intG(0,0,0,x,y) &=& 
2 y \trilog(1-x/y) + 2 x \trilog(1-y/x) 
+[3(x-y) - x \lnbar(x) + y \lnbar(y)] \dilog(1-x/y)
\nonumber \\ &&
-(x/3) \lnbar^3(x) + (y/6) \lnbar^3(y) 
+ x \lnbar^2(x) \lnbar(y)
- (x/2) \lnbar(x) \lnbar^2(y)
\nonumber \\ &&
-3 x \lnbar(x) \lnbar(y)
+ [3(x-y)/2] \lnbar^2(y)
+ (11x/2) \lnbar(x)
+ (11y/2) \lnbar(y)
\nonumber \\ &&
- (15/2 + 4 \zetathree/3) (x+y)
,
\label{eq:G000xy}
\\
\intG(0,0,x,0,y) &=&
(x + y) \bigl [-2 \trilog(1-x/y) - 2 \trilog(1-y/x) 
- (1/6) \lnbar^3(y) 
+ (1/3) \lnbar^3(x)
\nonumber \\ &&
+ 8 \zetathree/3 -\pi^2/6 - 15/2 \bigr ]
+ \left [2 (x - y) + (x + y) \ln(x/y) \right ] \dilog(1-x/y) 
\nonumber \\ &&	
+ (x/2 + y) \lnbar(x) \lnbar(y) \ln(y/x)
- (x/2) \lnbar^2(x)
+ (x-3y/2) \lnbar^2(y)
\nonumber \\ &&
- 2 x \lnbar(x) \lnbar(y)
+ [(33x + \pi^2 y)/6] \lnbar(x)
+ [(33y + \pi^2 x)/6] \lnbar(y)
,
\label{eq:G00x0y}
\\
\intG(x,0,0,0,y) &=& 
(y-x) \bigl \{ \trilog(1-x/y) + \trilog(1-y/x) 
+ \left [\lnbar(y) - 2 \right ] \dilog(1-x/y) 
\nonumber \\ &&
-(1/6) \lnbar^3(x) 
+ (1/3) \lnbar^3(y) 
- (1/2) \lnbar(x) \lnbar^2(y) 
- (\pi^2/6) \lnbar(x) 
- \zetathree/3 \bigr \} 
\nonumber \\ &&
+ (y/2) \lnbar^2(x) \lnbar(y)
- x \lnbar^2(x)
- 2 x \lnbar(x) \lnbar(y)
+ (x-3y/2) \lnbar^2(y)
\nonumber \\ &&
+ 4 x \lnbar(x)
+ (33/6 + \pi^2/6) y \lnbar(y)
- (7 + \pi^2) x/3
- (15/2 + \pi^2/6) y
,
\label{eq:Gx000y}
\\
\intG(x,0,x,0,y) &=&
-2 y \trilog(1-x/y) -2 y \trilog(1-y/x) 
+[3 x - 3 y + (2y-x) \lnbar(x) 
\nonumber \\ &&
- y \lnbar(y)] \dilog(1-x/y) 
+ (y/3) \lnbar^3(x)
- (y/6) \lnbar^3(y)
+ (x-y) \lnbar^2(x) \lnbar(y)
\nonumber \\ &&
+ (y-x/2) \lnbar(x) \lnbar^2(y)
- 2 x \lnbar^2(x)
- 3 x \lnbar(x) \lnbar(y)
+ [3(x-y)/2] \lnbar^2(y)
\nonumber \\ &&
+ (19 x/2) \lnbar(x)
+ (11 y/2) \lnbar(y)
-59 x/6 - 15 y/2+ 8 \zetathree y/3 
,
\label{eq:Gx0x0y}
\\
\intG(x,0,y,0,y) &=&
-[2(x-y)^2/x] \trilog(1-x/y) + 2 (x-y) [2 - \lnbar(y)] \dilog(1-x/y)
\nonumber
\\ && + [(y-2x)/3] \lnbar^3(y) + [x \lnbar(x) + 2 x - 3 y] \lnbar^2(y)
+ 4 x \lnbar(x) [1-\lnbar(y)] 
\nonumber
\\ && + 11 y \lnbar(y) - 7x/3 - 15 y + 
2 (1+y/x)(3y-x) \zeta_3/3 ,
\label{eq:Gx0y0y}
\\
\intG(0,x,x,y,y) &=&
(\sqrt{x} - \sqrt{y})^2 
\Bigl [ 2 \trilog(1-x/y) + 2 \trilog(1-y/x) + 
2 \ln(y/x) \dilog(1-x/y) 
\nonumber \\ &&
- (1/3) \lnbar^3(x) \Bigr ]
+ \sqrt{x y} \Bigl [
-32 \trilog(-\sqrt{x/y}) + 16 \ln(x/y) \dilog(-\sqrt{x/y})
\nonumber \\ &&
+ 4 \ln^2(x/y) \ln(1 + \sqrt{x/y}) 
+ 2 \lnbar(x) \lnbar(y) \ln(y/x) 
- (2/3) \lnbar^3(y) 
+ 4 \zetathree 
\Bigr ]
\nonumber \\ &&
+(x + y) [(2/3) \lnbar^3(y) -15 - 14 \zetathree/3]
+ (2x+y) \lnbar(x) \lnbar(y) \ln(x/y)
\nonumber \\ &&
- 3 x \lnbar^2(x) 
- 3 y \lnbar^2(y)
+ 11 x \lnbar(x)
+ 11 y \lnbar(y)
,
\label{eq:G0xxyy}
\\
G(x,0,x,y,y) &=& -F(x,0,y,y) + [2 - \lnbar(x)] I(x,y,y) + x \lnbar(x)/4 + 2 y \lnbar(y)
\nonumber \\ &&
- 11 x/12 - 14 y/3,
\label{eq:Gx0xyy}
\\
G(x,0,x,x,y) &=& -F(x,0,x,y) + [2 - \lnbar(x)] I(x,x,y) + 5 x \lnbar(x)/4 + y \lnbar(y)
\nonumber \\ &&
- 13x/4 - 7y/3,
\label{eq:Gx0xxy}
\\
G(y,x,x,x,x) &=& (4-y/x) \Fbar (y,0,x,x) + [8 - y/x - 2 \lnbar(x)] I(x,x,y)
\nonumber \\ &&
+ y [-8/3 - 2 \lnbar(x) - \lnbar(y) + 4 \lnbar(x) \lnbar(y) - \lnbar^2(x) \lnbar(y)]
\nonumber \\ &&
+ x [26/3 - 16 \lnbar(x) + 6 \lnbar^2(x)]
+ (y^2/4x) [17/3 - 3 \lnbar(y)] + (8 x - 2 y) \zeta_3,
\\
G(x,x,y,x,y) &=&
(y/x - 1) F(x, x, y, y) + y (y- 4 x) F(y, 0, x, x)/2 x^2
\nonumber \\ &&
+  [8 x^2 - 4 x y + y^2 + 2 x (y -2x) \lnbar(x) - 2 x y \lnbar(y)] I(x, x, y)/2 x^2 
\nonumber \\ &&
+ x [27/4  + (47/4) \lnbar(x) + 6 \lnbar^2(x) -\lnbar^3(x)]
\nonumber \\ &&
+ y [-217/12 + (47/4) \lnbar(x) + 7 \lnbar(y) - 3 \lnbar^2(x) - 4 \lnbar(x) \lnbar(y) 
+ \lnbar^2(x) \lnbar(y)] 
\nonumber \\ &&
+ (y^2/x) [29/6 - 4 \lnbar(x) - (11/2) \lnbar(y) + 4 \lnbar(x) \lnbar(y) + \lnbar^2(y)
- \lnbar(x) \lnbar^2(y)]
\nonumber \\ &&
+ (y^3/8 x^2) [-17/3 +3 \lnbar(y)]
+ (8y - 2 y^2/x) \zeta_3,
\\
\intG(x,0,0,y,y) &=& 
\sqrt{x^2 - 4 x y} \Bigl \{
2 \trilog(-k) + 4 \trilog(k/[1+k]) 
+ \dilog(-k) [2 \lnbar(y) - 4] 
\nonumber \\ &&
+ (1/12) \ln^3(k) 
 + \ln^2(k) [\lnbar(x) + \lnbar(y) - 4]/4
- \ln(k) [(1/4) \ln^2(x/y) + \pi^2/6]
\nonumber \\ &&
- (1/12) \ln^3(x/y) + \pi^2 (\lnbar(x) - 2)/6 - 2 \zetathree \Bigr \}
+ (y/3 - x/6) \lnbar^3(y) 
\nonumber \\ &&
+ (x/2) \lnbar^2(x) \lnbar(y) 
+(x-3y)\lnbar^2(y) - 2 x \lnbar(x)\lnbar(y) 
-x \lnbar^2(x) + 4 x \lnbar(x) 
\nonumber \\ &&
+ [\pi^2 x/6  + 11 y] \lnbar(y) 
+ 4 (x-2y)\zetathree/3 
- (7 + \pi^2) x/3 - 15 y  ,
\label{eq:Gx00yy}
\eeq
where $k$ in the last equation was given in eq.~(\ref{eq:defk}). 
Equations (\ref{eq:Gx000y}) and (\ref{eq:Gx00yy})
are equivalent to the results already obtained in
eqs.~(64) and (81) of ref.~\cite{Bytev:2009kb}.
The equivalent of eq.~(\ref{eq:Gx0y0y}) has also been obtained in terms of harmonic polylogarithms in ref.~\cite{Grigo:2012ji}.

In addition to the analytical cases, we find various identities that can be obtained
by requiring the absence of pole singularities in the derivatives of the basis integrals
for special values of the input squared masses.
For example, the following identities allow for
all remaining cases of $\intG$ with first argument vanishing to be written in terms of 
integral functions with fewer propagators:
\beq
G(0,u,v,y,z) &=& \Bigl \{
v \intF(v,u,y,z) - u \intF(u,v,y,z) 
 + [A(v) - A(u)]\intI(0,y,z)
\nonumber \\ &&  
 + [u A(u) - v A(v)]/4
\Bigr \}/(u-v) + 
\nonumber \\ &&
+ \Bigl \{
z \intF(z,y,u,v) - y \intF(y,z,u,v) 
 + [A(z) - A(y)]\intI(0,u,v) 
\nonumber \\ &&
 + [y A(y) - z A(z)]/4
\Bigr \}/(y-z)
- 2 (u+v+y+z)/3
,
\label{eq:G0uvyz}
\\
G(0,u,u,y,z) &=& 2 \Bigl \{
z \intF(z,y,u,u) 
- y \intF(y,z,u,u)
+ [A(z) - A(y)] \intI(0,u,u) 
\nonumber \\ &&
+ [y A(y) - z A(z)]/4 \Bigr \}/(y-z)
- \intF(u,u,y,z) - \lnbar(u) \intI(0,y,z)
\nonumber \\ &&
-A(y) - A(z) +  A(u)/4 
,
\label{eq:G0uuyz}
\eeq
This is useful because we find that when the first argument of $\intG$ 
vanishes, it tends to be especially sensitive to non-negligible numerical error from 
the Runge-Kutta integration described in sections \ref{sec:diffeq} 
and \ref{sec:3VIL} below, 
but we can always replace that value by the results of one of 
eqs.~(\ref{eq:G000xy}), (\ref{eq:G00x0y}), 
(\ref{eq:G0xxyy}), (\ref{eq:G0uvyz}), or (\ref{eq:G0uuyz}).

Another special identity is:
\beq
G(x,u,v,y,z) 
\bigl 
|_{u = (\sqrt{x} - \sqrt{v})^2} 
\bigr. 
&=&
\left (r - 1 \right ) F(u,v,y,z)
- r F(v,u,y,z)
\nonumber \\ &&
+ \left [1 - r A(v)/v + (r - 1) A(u)/u \right ] I(x,y,z) 
\nonumber \\ &&
+ (r/4) [A(v) +v] + [(1-r)/4][A(u) + u] + A(y) + A(z) 
\nonumber \\ &&
+ (19x - 41u -41v - 32y -32z)/24 ,
\eeq
where $r = \sqrt{v/x}$. In the special case $u=0$, this reduces to
\beq
G(x,0,x,y,z) &=& -F(x,0,y,z) + [1 - A(x)/x] I(x,y,z) +  A(x)/4 + A(y) + A(z)
\nonumber \\ &&
- (2x + 4y +4z)/3,
\eeq
which in turn has the fully analytic (in terms of ordinary polylogarithms) special cases
of eqs.~(\ref{eq:Gx0x0y}), (\ref{eq:Gx0xyy}), and (\ref{eq:Gx0xxy}).
Also, the 
following 4-propagator 
integral identity provides a useful check 
when one of the squared mass arguments vanishes:
\beq
0 &=& u (u-y-z) \intF(u,0,y,z) + y (y-u-z) \intF (y,0,u,z) + z (z-u-y) \intF(z,0,u,y)
\nonumber \\ &&
+ \lambda(u,y,z) \intI(u,y,z)
+ 2 A(u) A(y) A(z) - 2 u A(y) A(z) 
- 2 y A(u) A(z) - 2 z A(u) A(y)
\nonumber \\ &&
+ (3 u - 4 y - 4 z) (u - y - z) A(u)/4
+ (3 y - 4 u - 4 z) (y - u - z) A(y)/4
\nonumber \\ &&
+ (3 z - 4 u - 4 y) (z - u - y) A(z)/4
\nonumber \\ &&
+2 (u^2 y + u^2 z + y^2 u + y^2 z + z^2 u + z^2 y
 -u^3 - y^3 - z^3 )/3 .
\label{eq:Fu0yzcheck}
\eeq
These identities can be useful in reducing analytical expressions before numerical evaluation.

Finally, in all cases with two non-zero 
squared mass scales $x,y$, it is possible to find series
with expansion parameters including $y/x$, $x/y$, $(1-y/x)$, and sometimes 
$(1-4y/x)$, $(1-4x/y)$, $(1-9y/x)$, and/or $(1-9x/y)$,
so that the union of the overlapping regions of convergence cover all 
real positive $x,y$. 
In the code {\tt 3VIL} described below, we have incorporated such series results for 
all of the cases with at least one 0 squared mass argument and 
two other distinct squared masses, namely:
$H(0,0,0,0,x,y)$,
$H(0,0,x,y,0,0)$,
$H(0,0,0,y,x,x)$,
$H(0,0,y,0,x,x)$,
$H(0,0,x,x,0,y)$, 
$H(0,0,x,y,0,x)$,
$H(0,0,x,x,x,y)$,
$H(0,0,y,x,x,x)$,
$H(0,0,x,y,x,x)$, 
$H(0,0,x,x,y,y)$,
$H(0,0,x,y,x,y)$,  
$H(0,x,x,x,0,y)$,
$H(0,x,y,y,0,x)$,
$H(0,x,x,y,0,y)$,
$H(0,x,x,x,x,y)$, 
$H(0,x,x,x,y,x)$,
$H(0,x,x,x,y,y)$,
$H(0,x,y,x,x,y)$,
$H(0,x,x,y,y,y)$,
$H(0,x,y,y,x,x)$,
and permutations of them, together with the all of the 
subordinate 4-propagator and 5-propagator integrals of these that are 
not already given above analytically, namely:
$G(x,0,0,x,y)$, 
$G(y,0,x,x,x)$,
$G(x,0,y,x,x)$, 
$G(x,0,y,x,y)$,
$G(x,y,y,x,x)$,
$G(x,x,x,x,y)$,
$G(y,x,x,x,y)$,
$\Fbar(0,x,x,y)$, 
$\Fbar(x,x,x,y)$,
and 
$\Fbar(y,x,x,x)$. 
The coefficients of the terms
in the series expansions are implemented as pre-computed numerical values.
 
%%%%%%%%%%%%%%%%%%%%%%%%%%%%%%%%%%%%%%%%%%%%%%%%%%%%%%%%%%%%%%%
\section{Differential equations for numerical evaluation\label{sec:diffeq}}
\setcounter{equation}{0} 
\setcounter{figure}{0}
\setcounter{table}{0}
\setcounter{footnote}{1}

In this section we describe the differential equations method used for finding 
the 3-loop basis integrals in the case of generic squared mass arguments. 
The equations described below are implemented in the software package {\tt 3VIL},
as described in the following section.

For a given master tetrahedral topology corresponding to a basis integral 
\beq
H(u,v,w,x,y,z),
\label{eq:Huvwxyz}
\eeq 
the list of subordinate 3-loop basis integrals $\intG$ obtained by 
removing one propagator is:
\beq
&&\intG(w,u,z,v,y),\>\> \intG(x,u,v,y,z),\>\> \intG(u,v,x,w,z),\\ 
&&\intG(y,v,w,x,z),\>\>  \intG(v,u,x,w,y),\>\>  \intG(z,u,w,x,y).
\eeq
The list of subordinate $\Fbar$ integrals obtained by removing a second propagator is
\beq
&&\Fbar(w,u,x,y),\>\>  \Fbar(w,v,x,z),\>\>  \Fbar(x,u,w,y),\>\>  \Fbar(x,v,w,z),\\
&&\Fbar(u,v,y,z),\>\>  \Fbar(u,w,x,y),\>\>  \Fbar(y,u,v,z),\>\>  \Fbar(y,u,w,x),\\
&&\Fbar(v,u,y,z),\>\>  \Fbar(v,w,x,z),\>\>  \Fbar(z,u,v,y),\>\>  \Fbar(z,v,w,x).
\label{eq:Fbarlist}
\eeq
Also, there are associated 
2-loop basis integrals, obtained by removing from $H(u,v,w,x,y,z)$ 
any three propagators forming a complete loop:
\beq
&&\intI(u,v,x),\>\> \intI(x,y,z),\>\> \intI(u,x,y),\>\> \intI(v,x,z),\>\>
\intI(u,w,z),\>\> \intI(v,w,y), \\
&&\intI(u,w,y),\>\> \intI(v,w,z),\>\> \intI(v,y,z),\>\> \intI(w,x,y),\>\> 
\intI(u,v,z),\>\> \intI(u,w,x),\\
&&\intI(u,y,z),\>\> \intI(w,x,z),\>\> \intI(u,v,y),\>\> \intI(v,w,x).
\phantom{xxx}
\label{eq:Ilist}
\eeq
Although the $\intI$ functions 
are known analytically in terms of dilogarithms, in practice it is  
more efficient to treat them as dependent variables 
and solve for them simultaneously with the 3-loop basis functions. 

We now introduce a dimensionless 
independent variable $t$, and an arbitrary\footnote{In principle, 
the results should not
depend on the choice of $a$. By default, {\tt 3VIL} chooses 
$a= 2\,|\mbox{Max}(u,v,w,x,y,z)|$, which we find avoids some numerical
complications, with some exceptions noted below which require 
a different choice. As an option, $a$ can be specified at run time.
Changing $a$ allows a check on the numerical errors.} 
reference squared mass $a$, and define the quantities
\beq
U &=& a + t (u-a),
\qquad\quad
V \>=\> a + t (v-a),
\qquad\quad
W \>=\> a + t (w-a),
\nonumber \\
X &=& a + t (x-a),
\qquad\quad
Y \>=\> a + t (y-a),
\qquad\quad
Z \>=\> a + t (z-a).
\eeq
Now consider the 3-loop and 2-loop 
basis integrals, generically denoted $f_i$, as functions of arguments 
$(U,V,W,X,Y,Z)$, or equivalently
as functions of $u,v,w,x,y,z$ and $t$. These functions satisfy coupled
first-order differential equations of the general form:
\beq
\frac{d f_i}{dt} = \sum_j c_{ij} f_j + c_i.
\label{eq:diffeqs}
\eeq
Here, the $c_{ij}$ are ratios of polynomials in the squared masses and 
$t$ and, in the case where $f_j$ is an $\intI$ function, also linear 
functions of the logarithms $\lnbar(U), \lnbar(V)$, etc. The $c_{i}$ are 
up to cubic functions of the logarithms when $i$ is a 3-loop integral, 
and quadratic functions of the logarithms when $f_i$ is an $\intI$ 
integral. The differential equations are given explicitly below. These coupled 
differential equations in $t$ can be solved numerically by Runge-Kutta, 
using appropriate boundary conditions. At $t=0$, all of the propagator 
squared masses are equal to $a$, while at the endpoint of the 
integration $t=1$ we have $(U,V,W,X,Y,Z) = (u,v,w,x,y,z)$ equal to the 
desired values.

We now provide the derivatives of the basis integrals with respect to $t$. 
It is convenient to first define some auxiliary functions, in addition to
the functions $\lambda$ and $\psi$ defined in eqs.~(\ref{eq:deflambda})
and (\ref{eq:defpsi})
respectively:
\beq
\kappa (x,y,z) &=& x^2 + y^2 + z^2 - x y - x z - y z,
\\
\Delta (w,x,y,z) &=& \lambda(x,y,z) + 2 w (x+y+z) - 3 w^2,
\\
\phi (w,x,y,z) &=& \psi(w,x,y,z)
+ 8 a (w + x - y - z) (w - x + y - z) (w - x - y + z) ,
\eeq
Then define:
\beq
r_{\pm}(x,y,z) &=& 
a \bigl [x+y+z -3 a \pm 2 \sqrt{\kappa(x,y,z)}
\bigr ]/\lambda(a-x,a-y,a-z)
,
\eeq
and, if $\Delta(w,x,y,z) \not= 0$,
\beq
r_4(w,x,y,z) &=& 
8 a (w + x - y - z) (w - x + y - z) (w - x - y + z)/\phi(w,x,y,z) ,
\label{eq:r4generic}
\eeq
while in the alternative,
\beq
r_4(w,x,y,z) &=& a/(a-w)\qquad\quad\mbox{[if $\Delta(w,x,y,z)=0]$}.
\eeq
Note that if $\Delta(w,x,y,z) \not=0$, one must be careful not to choose $a$ 
to be the specific value such that $\phi(w,x,y,z) = 0$; 
otherwise a singularity would
occur in eq.~(\ref{eq:r4generic}). These are the exceptions
referred to in the previous footnote. Our program {\tt 3VIL} automatically
ensures that $a$ is chosen appropriately.

Then we can write:
\beq
\frac{d}{dt} \intI(X,Y,Z) &=& 
c_{II}(x,y,z) \intI(X,Y,Z)  
+ c_{ILL}(x,y,z) \lnbar(X) \lnbar(Y) 
\nonumber \\ &&
+ c_{ILL}(x,z,y) \lnbar(X) \lnbar(Z)  
+ c_{ILL}(y,z,x) \lnbar(Y) \lnbar(Z) 
\nonumber \\ &&
+ c_{IL}(x,y,z) \lnbar(X)
+ c_{IL}(y,x,z) \lnbar(Y)
+ c_{IL}(z,x,y) \lnbar(Z)
+ c_I(x,y,z),\phantom{xx}
\label{eq:dIdt}
\eeq
where we suppress the $a$ and $t$ dependences when writing the arguments of
the coefficient functions. These are given by
\beq
c_{II}(x,y,z) &=& 
\frac{1}{2(t - p_+)} + \frac{1}{2(t - p_-)},
\\
c_{ILL}(x,y,z) &=& 
\frac{c_{ILL+}}{t - p_+} + \frac{c_{ILL-}}{t - p_-} ,
\\
c_{IL}(x,y,z) &=& 
a-x + \frac{c_{IL+}}{t - p_+} + \frac{c_{IL-}}{t - p_-} ,
\\
c_{I}(x,y,z) &=& 
2x+2y+2z-6a + \frac{c_{I+}}{t - p_+} + \frac{c_{I-}}{t - p_-} ,
\eeq
with simple poles at 
\beq
p_\pm &=& r_\pm(x,y,z),
\label{eq:defp1}
\eeq
and coefficients:
\beq
c_{ILL\pm} &=& [a + (x+y-z-a) p_\pm]/4 ,
\\
c_{IL\pm} &=& (a-x) p_\pm - a,
\\
c_{I\pm} &=& 5 [3 a +(x+y+z- 3a) p_\pm]/4.
\label{eq:defcIpm}
\eeq
In {\tt 3VIL}, the $t$-independent coefficients appearing in 
eqs.~(\ref{eq:dIdt})-(\ref{eq:defcIpm}) and similar equations below
are computed only once, before the Runge-Kutta running begins.

Similarly, we find:
\beq
\frac{d}{dt} \Fbar(U,V,Y,Z) &=& 
c_{FF1}(u,v,y,z) \intF(U,V,Y,Z)  
+ c_{FF2}(u,v,y,z) \intF(V,U,Y,Z)
\nonumber \\ &&
+ c_{FF2}(u,y,v,z) \intF(Y,U,V,Z)
+ c_{FF2}(u,z,v,y) \intF(Z,U,V,Y)
\nonumber \\ &&
+ c_{FLLL1}(u,v,y,z) \lnbar(V) \lnbar(Y) \lnbar(Z)
+ c_{FLLL2}(u,v,y,z) \lnbar(U) \lnbar(V) \lnbar(Y)
\nonumber \\ &&
+ c_{FLLL2}(u,v,z,y) \lnbar(U) \lnbar(V) \lnbar(Z)
+ c_{FLLL2}(u,y,z,v) \lnbar(U) \lnbar(Y) \lnbar(Z)
\nonumber \\ &&
+ c_{FLL1}(u,v,y,z) \lnbar(V) \lnbar(Y)
+ c_{FLL1}(u,v,z,y) \lnbar(V) \lnbar(Z)
\nonumber \\ &&  
+ c_{FLL1}(u,y,z,v) \lnbar(Y) \lnbar(Z)
+ c_{FLL2}(u,v,y,z) \lnbar(U) \lnbar(V)
\nonumber \\ && 
+ c_{FLL2}(u,y,v,z) \lnbar(U) \lnbar(Y)
+ c_{FLL2}(u,z,v,y) \lnbar(U) \lnbar(Z)
\nonumber \\ && 
+ c_{FIL}(u,v,y,z) \lnbar(U) \intI(V,Y,Z)
+ c_{FI}(u,v,y,z) \intI(V,Y,Z)
\nonumber \\ && 
+ c_{FL1}(u,v,y,z) \lnbar(U)
+ c_{FL2}(u,v,y,z) \lnbar(V)
+ c_{FL2}(u,y,v,z) \lnbar(Y)
\nonumber \\ && 
+ c_{FL2}(u,z,v,y) \lnbar(Z)
+ c_F(u,v,y,z),\phantom{vv}
\eeq
Note that the right side contains $\intF$ functions, which 
in the {\tt 3VIL} code are expressed in terms
of $\Fbar$ functions using eq.~(\ref{eq:defFbar}). 
The coefficient functions on the right side again can be written as 
sums over simple poles in $t$:
\beq
c_{FF1}(u,v,y,z) &=& \frac{3}{4 t} + \frac{1}{4(t-p_3)} ,
\label{eq:defcFF1}
\\
c_{FF2}(u,v,y,z) &=& -\frac{1}{4t} + \frac{c_{FF22}}{t-p_2} +
 \frac{c_{FF23}}{t-p_3} ,
\\
c_{FLLL1}(u,v,y,z) &=& \frac{3a}{4t} + \frac{c_{FLLL12}}{t-p_2} +
 \frac{c_{FLLL13}}{t-p_3} ,
\\
c_{FLLL2}(u,v,y,z) &=& -\frac{a}{4t} +
 \frac{c_{FLLL23}}{t-p_3} +  \frac{c_{FLLL24}}{t-p_4} + \frac{c_{FLLL25}}{t-p_5} ,
\\
c_{FLL1}(u,v,y,z) &=& -\frac{a}{t} + \frac{c_{FLL12}}{t-p_2} +
 \frac{c_{FLL13}}{t-p_3} ,
\\
c_{FLL2}(u,v,y,z) &=& \frac{a}{t} + 
 \frac{c_{FLL23}}{t-p_3} +  \frac{c_{FLL24}}{t-p_4} + \frac{c_{FLL25}}{t-p_5} ,
\\
c_{FIL}(u,v,y,z) &=& \frac{1}{2(t-p_4)} + \frac{1}{2(t-p_5)},
\\
c_{FI}(u,v,y,z) &=& \frac{1}{t-p_2} ,
\\
c_{FL1}(u,v,y,z) &=& u-a - \frac{63 a}{16t} + 
 \frac{c_{FL13}}{t-p_3} +  \frac{c_{FL14}}{t-p_4} + \frac{c_{FL15}}{t-p_5} ,
\\
c_{FL2}(u,v,y,z) &=& a-v + \frac{21 a}{16 t} + 
 \frac{c_{FL22}}{t-p_2} +  \frac{c_{FL23}}{t-p_3} ,
\\
c_{F}(u,v,y,z) &=& -\frac{13}{4}a - \frac{11}{4}u + 2 v + 2y + 2 z  + \frac{c_{F2}}{t-p_2} +  \frac{c_{F3}}{t-p_3} ,
\label{eq:defcF}
\eeq
where the coefficients $c_{FF22}$ etc. on the right side are independent of
$t$, and there are simple poles in $t$ at
\beq
p_1 &=& 0,
\\
p_2 &=& a/(a-u),
\\
p_3 &=& r_4(u,v,y,z),
\\
p_{4,5} &=& r_\pm (v,y,z) .
\eeq
Note that there are always poles at $t=0$. If a squared mass argument 
vanishes, then there will also be a pole at $t=1$. The explicit forms 
for some of the $t$-independent coefficients on the right sides of 
eqs.~(\ref{eq:defcFF1})-(\ref{eq:defcF}) are somewhat complicated, so 
they are relegated to an ancillary electronic file called {\tt 
dFbardtcoeffs.txt}, which is included with the arXiv submission for this 
paper. There are two separate forms for these coefficients, depending on 
whether $\Delta(u,v,y,z)$ is zero or non-zero.

The differential equations for the $\intG$ functions have the form:
\beq
\frac{d}{dt} \intG(W,U,Z,V,Y) &=& 
c_{GG}(w,u,z,v,y) \intG(W,U,Z,V,Y) 
\nonumber \\ && 
+ c_{GF}(w,u,z,v,y) [\intF(U,V,Y,Z) + \lnbar(U) \intI(V,W,Y)]
\nonumber \\ && 
+ c_{GF}(w,z,u,v,y) [\intF(Z,U,V,Y) + \lnbar(Z) \intI(V,W,Y)]
\nonumber \\ && 
+ c_{GF}(w,v,y,u,z) [\intF(V,U,Y,Z) + \lnbar(V) \intI(U,W,Z)]
\nonumber \\ && 
+ c_{GF}(w,y,v,u,z) [\intF(Y,U,V,Z) + \lnbar(Y) \intI(U,W,Z)]
\nonumber \\ &&
+ c_{GI}(w,u,z,v,y) \intI(U,W,Z)
+ c_{GI}(w,v,y,u,z) \intI(V,W,Y)
\nonumber \\ &&
+ c_{GL}(w,u,z,v,y) U\lnbar(U)
+ c_{GL}(w,z,u,v,y) Z\lnbar(Z)
\nonumber \\ &&
+ c_{GL}(w,v,y,u,z) V\lnbar(V)
+ c_{GL}(w,y,v,u,z) Y\lnbar(Y)
\nonumber \\ &&
+ c_{G}(w,u,z,v,y) 
\eeq
where again the $\intF$ functions on the right side are 
re-expressed in terms of
$\Fbar$ functions in the {\tt 3VIL} code using eq.~(\ref{eq:defFbar}). 
The coefficient functions are:
\beq
c_{GG}(w,u,z,v,y) &=&
-\frac{1}{t-p_1}
+ \frac{1}{2} \left [\frac{1}{t-p_2}
+\frac{1}{t-p_3}
+\frac{1}{t-p_4}
+\frac{1}{t-p_5}\right ]
,
\label{eq:cGG}
\\
c_{GF}(w,u,z,v,y) &=&
\frac{c_{GF1}}{t-p_1}
+\frac{c_{GF4}}{t-p_4}
+\frac{c_{GF5}}{t-p_5}
,
\label{eq:cGF}
\\
c_{GI}(w,u,z,v,y) &=&
\frac{1}{t-p_1}
-\frac{1}{t-p_2}
-\frac{1}{t-p_3}
,
\\
c_{GL}(w,u,z,v,y) &=& 
-\frac{1}{2(t-p_2)}
-\frac{1}{2(t-p_3)}
-\frac{1}{4} c_{GF}(w,u,z,v,y)
,
\\
c_{G}(w,u,z,v,y) &=&
c_{G0} 
+\frac{c_{G1}}{t-p_1}
+\frac{c_{G2}}{t-p_2}
+\frac{c_{G3}}{t-p_3}
+\frac{c_{G4}}{t-p_4}
+\frac{c_{G5}}{t-p_5}
,
\label{eq:cG}
\eeq
with simple poles at
\beq
p_1 &=& a/(a-w),
\\
p_{2,3} &=& r_\pm(y,v,w),
\\
p_{4,5} &=& r_\pm(u,z,w).
\eeq
The coefficients in eq.~(\ref{eq:cG}) are given by
\beq
c_{G0} &=& -11 a - w + 3 u + 3 v + 3 y + 3 z,
\\
c_{G1} &=& 11 a (4 w - u - v - y - z)/12(a - w),
\\
c_{G2,3} &=& 175 a/48 + p_{2,3} (-175 a - 19 w + 56 u + 56 z + 41 v + 41 y)/48,
\\
c_{G4,5} &=& 175 a/48 + p_{4,5} (-175 a - 19 w + 56 v + 56 y + 41 u + 41 z)/48,
\eeq  
and those in eq.~(\ref{eq:cGF}) are given by, if $u\not=z$:
\beq
c_{GF1} &=& (w-u)/(u-z),
\\
c_{GF4,5} &=& \bigl [u - w \pm \sqrt{\kappa(u,w,z)} \bigr ]/2 (u - z),
\eeq
while if $u=z$ they are:
\beq
c_{GF1} &=& 0,
\\
c_{GF4,5} &=& \pm{\rm sign}(w-u)/4,
\eeq
with ${\rm sign}(x) = x/|x|$ if $x \not=0$, and ${\rm sign}(0)\equiv 0$.

Finally, the differential equation for $\intH$ is:
\beq
\frac{d}{dt} \intH(U,V,W,X,Y,Z) &=& 
c_{HG}(u,v,w,x,y,z) \Bigl \{ 
\intG(X,U,V,Y,Z) - \intF(X,V,W,Z) 
\nonumber \\ &&
- \intF(X,U,W,Y) 
+ [2 - \lnbar(X)] [\intI(U,W,Z) + \intI(V,W,Y)]
\nonumber \\ &&
+ W \lnbar(W) 
+ X \lnbar(X)/2 - 5 X/2 - 7 W/3 \Bigr \}
+ \mbox{(5 permutations)}\phantom{xx}
\nonumber \\ &&
+ c_{H}(u,v,w,x,y,z),
\eeq
where the ``(5 permutations)" of squared masses $(u,v,w,x,y,z)$
are determined by the tetrahedral symmetry of
Figure \ref{fig:HGFE}, and are given by
$(u, z, x, w, y, v)$ 
and
$(u, w, v, z, y, x)$
and
$(u, x, z, v, y, w)$ 
and
$(w, v, u, y, x, z)$ 
and 
$(x, v, y, u, w, z)$.
The coefficient functions have the forms:
\beq
c_{HG}(u,v,w,x,y,z) &=& \frac{c_{HGn}(u,v,w,x,y,z)}{
c_{Hd1}(u,v,w,x,y,z) c_{Hd2}(u,v,x) c_{Hd2}(x,y,z)}, \phantom{xx}
\label{eq:cHG}
\\
c_{H}(u,v,w,x,y,z) &=&  \frac{c_{Hn}(u,v,w,x,y,z)}{
c_{Hd1}(u,v,w,x,y,z)} \zeta_3
\eeq
where $c_{HGn}$, $c_{Hd1}$, $c_{Hd2}$, and $c_{Hn}$ are polynomials
in $t$ of orders $4$, $3$, $2$, and $2$, respectively. The roots of
the quadratic polynomials
$c_{Hd2}(u,v,x)$ and $c_{Hd2}(x,y,z)$ are respectively
$t=r_\pm(u,v,x)$ and $t = r_\pm(x,y,z)$.
The cubic polynomial in $t$ appearing in these denominators is:
\beq
c_{Hd1}(u,v,w,x,y,z) &=& 
-2 a^3 
+ (6a -u-v-w-x-y-z) a^2 t 
+ [-6 a^2 + 2 a (u+v 
\nonumber \\ &&
\!\!\!\!\!\!\!\!\!\!\!\!\!\!\!\!\!\!\!\!\!\!\!\!\!\!\!\!\!\!\!\!\!\!\!\!\!\!\!\!
+w+x+y+z) 
+ u^2 + v^2 + w^2 + x^2 + y^2 + z^2 
-u v - u w - v w - u x
\nonumber \\ &&
\!\!\!\!\!\!\!\!\!\!\!\!\!\!\!\!\!\!\!\!\!\!\!\!\!\!\!\!\!\!\!\!\!\!\!\!\!\!\!\!
- v x - v y - w y - x y - u z - w z - x z - y z] a t^2
+ [ u v x - u w x - v w x - u v y 
\nonumber \\ &&
\!\!\!\!\!\!\!\!\!\!\!\!\!\!\!\!\!\!\!\!\!\!\!\!\!\!\!\!\!\!\!\!\!\!\!\!\!\!\!\!
- u w y + v w y - u x y - w x y - u v z 
+ u w z - v w z - v x z - w x z - u y z - v y z
\nonumber \\ &&
\!\!\!\!\!\!\!\!\!\!\!\!\!\!\!\!\!\!\!\!\!\!\!\!\!\!\!\!\!\!\!\!\!\!\!\!\!\!\!\!
+ x y z
+ u^2 y + u y^2 + v^2 z + v z^2 + w^2 x + w x^2 
+ a ( u v + u w + v w + u x + v x
\nonumber \\ &&
\!\!\!\!\!\!\!\!\!\!\!\!\!\!\!\!\!\!\!\!\!\!\!\!\!\!\!\!\!\!\!\!\!\!\!\!\!\!\!\! 
+ v y + w y + x y + u z + w z + x z + y z
-u^2 - v^2 - w^2 - x^2 - y^2  - z^2 )
\nonumber \\ &&
\!\!\!\!\!\!\!\!\!\!\!\!\!\!\!\!\!\!\!\!\!\!\!\!\!\!\!\!\!\!\!\!\!\!\!\!\!\!\!\!
- a^2 (u+v+w+x+y+z)
+ 2 a^3] t^3 .
\label{eq:cubicpoly}
\eeq
In terms of the roots 
$R_{1,2,3}$ of this cubic polynomial in $t$, the expression for $c_{H}$ can be rewritten in a very simple form:
\beq
c_{H}(u,v,w,x,y,z) &=& -2 \zeta_3 \left [ \frac{1}{t-R_1} + \frac{1}{t-R_2} + \frac{1}{t-R_3}
\right ] .
\eeq
Unfortunately, however, attempting to write $c_{HG}$ 
as a sum of simple poles leads to extremely complicated expressions 
related to the solutions of a cubic equations, 
with residue coefficients that are also singular in a variety of special cases
for the squared masses.
Therefore, we instead write:
\beq
c_{HG}(u,v,w,x,y,z) &=& \Bigl (\sum_{k=0}^{4} c_{HGn}^{(k)} \,t^k \Bigr ) /
\Bigl ( \sum_{k=0}^{7} c_{HGd}^{(k)} t^k \Bigr ),
\eeq
with coefficients $c_{HGn}^{(k)}$ and $c_{HGd}^{(k)}$ that are 
complicated polynomials in $u,v,w,x,y,z$. They are given in an ancillary 
file called {\tt cHG.txt}, both in the generic case and in all special 
cases involving degenerate masses in which simplification occurs because 
the numerator and denominator can be reduced by a common factor. This is 
important for the Runge-Kutta evaluation because it avoids spurious 
higher-order poles at (or near) $t=1$ when one or more squared masses 
vanishes (or is relatively small). All of the natural special cases 
involving one or more degenerate squared masses are identified and 
treated separately within {\tt cHG.txt}. The computer library {\tt 3VIL} 
automatically identifies and deals with these special cases. It should 
be noted that there are other special cases of squared mass arguments in 
which our expression for $c_{HG}$ has higher-order poles in $t$, but 
those are all unnatural in the sense that they require relationships 
between squared masses that are not degeneracies and not consequences of 
any possible symmetry of a quantum field theory. At, and near, such 
unnatural special points one should be aware that there may be some loss 
of numerical precision.

From the above results, we can now make a list of all of the poles in $t$ in the 
complete set of coupled differential equations. They consist of the 
union of the points: $0$, and $a/(a-x_i)$ for $x_i = u,v,w,x,y,z$, and 
$r_\pm(x_i,x_j,x_k)$ for every triplet of arguments of $\intI$ functions 
appearing in eq.~(\ref{eq:Ilist}), and $r_4(x_i,x_j,x_k,x_l)$ for every 
quartet of arguments of $\Fbar$ functions appearing in 
eq.~(\ref{eq:Fbarlist}), and the three roots $R_1$, $R_2$, and $R_3$ of 
the cubic equation (\ref{eq:cubicpoly}).

%%%%%%%%%%%%%%%%%%%%%%%%%%%%%%%%%%%%%%%%%%%%%%%%%%%%%%%%%%%%%%%
\section{Implementation %of the differential equations method 
in software: 3VIL 2.0 \label{sec:3VIL}}
\setcounter{equation}{0} 
\setcounter{figure}{0}
\setcounter{table}{0}
\setcounter{footnote}{1}

In this section, we describe version 2.0 of the software package {\tt 
3VIL}, available at \cite{3VILwebsites}, which takes inputs 
$u,v,w,x,y,z$ and the renormalization scale $Q$, and outputs the 
numerical values of all of the basis integrals listed above using either 
the analytic expressions from section \ref{sec:analytic}, or, when those 
do not apply, a simultaneous Runge-Kutta computation involving their 
coupled differential equations in $t$ found in the previous section.

The $t=0$ values are known from eqs.~(\ref{eq:Ixxx}),
(\ref{eq:Fbarxxxx}), (\ref{eq:Gxxxxx}), and (\ref{eq:Hxxxxxx}), 
and so in principle could serve as boundary conditions for the 
Runge-Kutta integration. However, there is a technical difficulty in 
that some of the coefficients $c_{ij}$ and $c_{i}$ have unavoidable poles at $t=0$, 
even though the basis integral functions are always well-defined and smooth there. 
Therefore, we instead choose to integrate starting from a small non-zero 
value of $t$. This is done by first analytically solving the coupled differential 
equations as power series expansions in $t$:
\beq
\intI (X,Y,Z) &=&  \intI(a,a,a) + \sum_{n \geq 1} t^n \intI^{(n)}(x,y,z;a) , 
\\
\Fbar (U,V,Y,Z) &=&  \Fbar(a,a,a,a) + \sum_{n \geq 1} t^n \Fbar^{(n)}(u,v,y,z;a) ,
\\
\intG (W,U,Z,V,Y) &=&  \intG(a,a,a,a,a) + \sum_{n \geq 1} t^n \intG^{(n)}(w,u,z,v,y;a) ,
\\
\intH (U,V,W,X,Y,Z) &=&  \intH(a,a,a,a,a,a) + \sum_{n \geq 1} t^n \intH^{(n)}(u,v,w,x,y,z;a)
.
\eeq
The leading order ($t^0$) terms can be read off immediately from 
eqs.~(\ref{eq:Ixxx}),
(\ref{eq:Fbarxxxx}), (\ref{eq:Gxxxxx}), and (\ref{eq:Hxxxxxx}), 
and the coefficients of $t^1$ are given by:
\beq
&&\intI^{(1)}(x,y,z;a) =
(3a - x - y - z) \left [ 
\frac{1}{2} - \sqrt{3} \Lstwo - \lnbar(a) + \frac{1}{2} \lnbar^2(a) \right ] ,
\\
&&\Fbar^{(1)}(u,v,y,z;a) =
(7 u - 2 z - 2 y - 2 v - a)/6 + 3 \sqrt{3} \Lstwo (u-a) 
+ \frac{7}{3} \zetathree (v+y+z-3u)
\nonumber \\ &&
\qquad \qquad
+ \bigl [ \sqrt{3} \Lstwo (v + y + z - 3 a) + 
(a - 7 u + 2 v + 2 y + 2 z)/4 \bigr ] \lnbar(a)
\nonumber \\ &&
\qquad \qquad
+ \frac{1}{2} (u-a) \lnbar^2(a) + \frac{1}{6} (3 a - v - y- z) \lnbar^3(a)
,
\\
&&\intG^{(1)}(w,u,z,v,y;a) =
19 a/3 + 5 w/3 - 2 u - 2 v - 2 y - 2 z
+ 2(\sqrt{3} \Lstwo + \zetathree)(u+v+y+z
\phantom{xxxx}
\nonumber \\ &&
\qquad\qquad
-w-3a)
+ \bigl [\sqrt{3} \Lstwo (6a-2w-u-v-y-z) + 5 (u+v+y+z-4a)/2\bigr ] 
\lnbar(a)
\nonumber \\ &&
\qquad\qquad
+ (5a-w-u-v-y-z) \lnbar^2(a) + [(2w+u+v+y+z-6a)/6] \lnbar^3(a)
,
\\
&&\intH^{(1)}(u,v,w,x,y,z;a) = \zetathree (6 a - u - v - w - x - y - z)/a
. 
\eeq
We have computed the remaining terms of these expansions up through 
order $t^8$. These results are provided in an ancillary files, called 
{\tt texpansions.txt}, provided with the arXiv sources for this paper. 
In {\tt 3VIL}, we use these expansions to initiate the Runge-Kutta 
running at a small non-zero value $t=t_{\rm in}$ with magnitude $0.013$, 
so that the associated numerical relative error is of the order 
$10^{-16}$, comparable to the round-off error for long double arithmetic.

\begin{figure}[!tb]
  \begin{minipage}[]{0.55\linewidth}
    \includegraphics[width=8.5cm,angle=0]{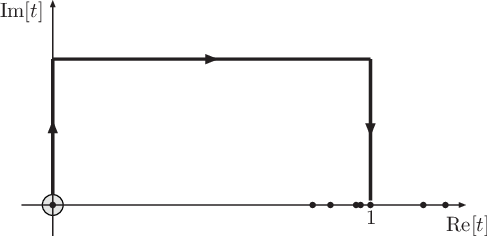}
  \end{minipage}
  \begin{minipage}[]{0.44\linewidth}
    \caption{\label{fig:contour} Schematic of the contour in the 
   upper-half complex $t$ plane for simultaneous evaluation of the basis integrals in 
   eqs.(\ref{eq:Huvwxyz})-(\ref{eq:Ilist}) using Runge-Kutta integration. 
   The integrals are initiated near the origin using an expansion for small $|t| = 0.013$, 
   and the contour finishes at $t=1$. The dots on the Re$[t]$
   axis represent poles in the coefficients in the differential equations.
   }
  \end{minipage}
\end{figure}

Another complication is that the coefficient functions $c_{ij}$ and 
$c_{i}$ also have poles at non-zero $t$. These poles always lie on the real 
$t$ axis; their locations were listed at the end of section \ref{sec:diffeq}.  
For many (but not all) choices of inputs $u,v,w,x,y,z$, one or more of the these 
poles will lie in the range of $t$ between 0 and 1 (for any choice of 
$a$). In order to avoid numerical problems when such poles are present 
with $0<t<1$, we promote $t$ to a complex variable, and integrate the 
coupled differential equations (\ref{eq:diffeqs}) in the upper\footnote{
A vacuum loop integral function of real squared mass arguments
can have an imaginary part if, 
and only if, one or more of 
the arguments is negative. 
For example, 
%$A(x) = x (\lnbar(|x|) - i \pi - 1)$ 
integral functions dependent on one mass scale $x$ are obtained by taking
$\lnbar(x) \rightarrow \lnbar(|x|) - i \pi$
for real negative $x$. 
More generally, approaching $t=1$ from above in the complex $t$ plane
provides the correct $m^2 - i\epsilon$ Feynman propagator 
prescription, and thus ensures that the
imaginary parts of the integral
functions will have the correct signs
when one or more squared masses is negative.} 
half complex $t$ plane 
along a contour that avoids 
the real $t$ axis, as shown in Figure 
\ref{fig:contour}.  
By default, the displacement of the contour in the Im$(t)$ direction
is 0.8, but this can be changed by the user at run time.
The initial point is chosen to be $t = i |t_{\rm in}|$ in this case, 
by default. In the 
nicer case of inputs $u,v,w,x,y,z$, and $a$ such that there is no pole 
in any of the coefficients $c_{ij}$ or $c_i$ for $0<t<1$, we save time 
and numerical accuracy by integrating the coupled differential equation 
directly along the real axis from $t = |t_{\rm in}|$ to $t= 1$.  
The user can also change the default value of the magnitude of the
starting point $|t_{\rm in}|$ from 0.013 to another value at run time.
%%%%%%%%%%%%%%%%%%%%%%%%%%%%%%%%%%%%%%%%%%%%%%
%%%%%%%%%%%%%%%%%%%%%%%%%%%%%%%%%%%%%%%%%%%%%%

The Runge-Kutta running is performed with the 6-stage, 5th-order 
Cash-Karp algorithm \cite{CashKarp} with automatic step-size adjustment. 
However, in some cases, the endpoint $t=1$ is also a pole of one or more 
of the coefficients $c_{ij}$ and $c_i$, even though all of the $\intI, 
\Fbar, \intG,$ and $\intH$ integrals are well-defined there. (For example, this 
occurs if any of $u,v,w,x,y,z$ vanishes.) In these 
cases, we need to use a somewhat unusual Runge-Kutta integration 
algorithm for the final step, such that there are no evaluations of 
coefficients at the final endpoint. We encountered a very similar 
problem in the case of {\tt TSIL}, and here we employ exactly the same 
solution as described there, involving a particular choice of 5-stage, 
4th-order Butcher coefficients.  The reader is referred to 
ref.~\cite{TSIL} for a more detailed description of this rather 
specialized Runge-Kutta strategy.

In version 2.0 it is possible to evaluate subsets of the basis 
integrals. There are two basic modes: (1) the EFG subset, consisting of 
a single $G$ function and all the subsidiary integrals needed for its 
evaluation, namely four $\Fbar$ functions and six $I$ functions, along 
with the single $E$ function that can be computed from these; and (2) 
the EF subset, consisting of one $E$ function and the subsidiary 
integrals needed for its evaluation, namely four $\Fbar$ functions and 
four $I$ functions. EFG subset integration is typically about 7 times 
faster than the full set of functions, for generic cases, with EF subset 
evaluation a further 25\% faster than EFG.

In special cases where the analytical values of one or more of the 
integrals is known, {\tt 3VIL} automatically replaces the values 
obtained by Runge-Kutta by the results of the analytical formulas [or 
reduction of $\intG(0,u,v,y,z)$ to $\Fbar$ and $\intI$ functions], using 
the results of section \ref{sec:analytic}. This is particularly useful 
because we find that the cases in which this is possible tend to be also 
cases in which the Runge-Kutta running is subject to relatively 
larger numerical errors.

Finally, for cases with three distinct non-zero mass scales, of the form:

~~~~~Case A: $(u,v,w,x,y,z) = (0,0,Y,Z,X,X)$

~~~~~Case B: $(u,v,w,x,y,z) = (0,X,X,Y,Z,Y)$

~~~~~Case C: $(u,v,w,x,y,z) = (0,X,Y,X,Z,Y)$

\noindent
and cases related to these by permutation, a special evaluation mode is 
used; this is new in version 2.0. In each of these cases, the 
Runge-Kutta running is carried out in terms of $Z$, with $X$ and $Y$ 
held fixed. The running starts from the known analytical values at $Z=0$, and 
proceeds either along the real $Z$-axis or in the complex $Z$ plane, 
depending on the singularity structure of the integrand coefficients. 
This approach is significantly faster and more accurate for these 
special cases, which often arise in practice, including in the 
evaluation of the Standard Model effective potential.

For illustration,\footnote{The code used to obtain the data in this 
figure is included with the {\tt 3VIL} distribution, as one example of 
how to use the software. Another provided sample user application 
program shows how to compute and extract all of the basis integrals for 
the case $(u, v, w, x, y, z) = (t, t, b, h, W, W)$ in the Standard 
Model, where particle names are used to represent squared masses.} we 
show in Figure \ref{fig:propaganda} the results for the integral $H$ for 
selected one-parameter families of arguments, parameterized by a single 
variable squared mass $0 \leq x \leq 1$. The other non-zero squared mass 
arguments and the renormalization scale $Q$ are chosen to be 1 in these 
examples. The values at the endpoints $x=0$ and $x=1$ are analytically 
known, and given in eqs.~(\ref{eq:H00000x})-(\ref{eq:Hxxxxxx}). Note 
that these integral functions vary smoothly with $x$, and tend to 
decrease as the squared mass arguments are increased.
%%%%%%%%%%%%%%%%%%%%%%%%%%%%%%%%%%%%%%%%%%%%%%%%%%%%%%%%%%%%%%%%
\begin{figure}[t]
\begin{minipage}[]{0.495\linewidth}
    \includegraphics[width=\linewidth,angle=0]{Ha.eps}
\end{minipage}
\begin{minipage}[]{0.495\linewidth}
    \includegraphics[width=\linewidth,angle=0]{Hb.eps}
\end{minipage}
\caption{\label{fig:propaganda} Numerical values of the integral $H$
for selected squared mass arguments, from top to bottom $H(x,x,x,x,x,1)$,
$H(x,x,1,1,x,x)$, $H(x,x,1,x,1,1)$, $H(x,1,1,1,x,1)$, and $H(x,1,1,1,1,1)$
(left panel), and $H(0,0,0,0,x,1)$, $H(0,0,0,x,x,1)$, $H(0,0,x,x,x,1)$,
$H(0,x,x,x,1,x)$, $H(0,1,x,x,x,1)$, $H(0,x,1,1,x,1)$, and $H(0,1,1,1,x,1)$ (right panel),
as a function of  $0\leq x\leq 1$, 
as computed by {\tt 3VIL} using the Runge-Kutta solution of the 
differential equations in $t$. 
In each case, the renormalization scale is $Q=1$.
At the endpoints $x=0$ and $x=1$, each of the values shown agrees 
with an analytic special case given in 
eqs.~(\ref{eq:H00000x})-(\ref{eq:Hxxxxxx}).}
\end{figure}
%%%%%%%%%%%%%%%%%%%%%%%%%%%%%%%%%%%%%%%%%%%%%%%%%%%%%%%%%%%%%%%%

We have also checked consistency of 
all of the other analytic special cases
for $I$, $E$, $F$, $\Fbar$, $G$, and $H$ 
functions in section \ref{sec:analytic}, compared to
the results obtained from Runge-Kutta integration of the differential equations in $t$.
The results reported to the user by {\tt 3VIL} are always 
the analytic ones, when they are available.

For input squared masses $u,v,w,x,y,z$ and renormalization scale $Q$, 
{\tt 3VIL} automatically evaluates simultaneously all of the basis 
functions $H$, $G$, $\Fbar$, and $I$, and the associated functions $E$ 
and $F$ and $I_\epsilon$, as well as the alternative basis bold 
functions (for those who may prefer them), 
${\bf E},$ ${\bf F},$ ${\bf G},$ and ${\bf H}$ in the conventions
and notation given in 
section III above. 
The latter are evaluated and stored as the 
coefficients of $\epsilon^{-n}$ for $n=0,1,2,3$ ($n=0,1$ only for 
${\bf H}$). Utilities are provided in {\tt 3VIL} for extracting the basis 
%%DGR
function values from the results struct after computation, for 
permuting results according to the tetrahedral symmetry group of $H$, for 
printing results, etc.

Although the integral functions are always real for non-negative squared 
mass arguments, they are computed and given as long double complex 
numbers. The magnitude of the imaginary part, which arises due to the 
Runge-Kutta integration off of the real axis in the complex $t$ plane, 
therefore can serve as a check of accuracy of the calculation, as it 
should vanish in the idealized case of no computational error. 
Integration off of the real axis is not always necessary, and is avoided 
by default when possible, but if desired it can be forced by the user, 
and the magnitude of the deviation of the contour from the real $t$ axis can 
be varied by the user, as a check. We find that the magnitude of the 
imaginary part computed by the Runge-Kutta method is often larger than 
the error in the real part (determined either by analytical evaluation 
when possible, or by varying the default characteristics of the 
integration), so we expect that the imaginary part is often a 
conservative error estimate.

For generic input parameters, the relative accuracy of the results is 
typically on the order of $10^{-9}$ or better, but 
it can be worse for difficult 
cases corresponding to pseudo-thresholds where some triplet of squared 
masses $(x,y,z)$ of propagators meeting at a vertex have a small 
magnitude of $|\sqrt{x} \pm \sqrt{y} \pm \sqrt{z}|$. Even in the worst 
cases of $H$ integrals with more than one such pseudo-threshold, the 
relative accuracy is typically about $10^{-4}$ or better, 
which should be good 
enough for practical applications at 3-loop order.
For generic input parameters, the total computation time by {\tt 3VIL}
for the simultaneous computation of all of the integrals is well 
under 1 second on modern hardware, but it can be somewhat more for 
the especially difficult cases. For analytical cases, the computation 
time is extremely short and relatively negligible.

The {\tt README.txt} file included with the
{\tt 3VIL} distribution available at \cite{3VILwebsites} provides
additional technical details regarding the numerical integration 
techniques employed, a complete description of the user application 
programming interface, and some sample user programs illustrating how to use
the library.

%%%%%%%%%%%%%%%%%%%%%%%%%%%%%%%%%%%%%%%%%%%%%%%%%%%%%%%%%%%%%%%%%%%%%%%%
\section{Outlook\label{sec:Outlook}}
\setcounter{equation}{0}
\setcounter{figure}{0}
\setcounter{table}{0}
\setcounter{footnote}{1}

In this paper, we have studied the basis functions for 3-loop vacuum Feynman integrals,
and provided results and a public open-source software package,
available at \cite{3VILwebsites},
to efficiently evaluate 
them. We plan to maintain, update, and improve the code package {\tt 3VIL} indefinitely, and welcome suggestions and bug reports.
 
One obvious application of these results 
is to the computation of the effective potential (or its derivatives) 
for a general theory,
and for the Standard Model in particular, at full 3-loop order. 
At present, the Standard Model effective potential is known at 2-loop order
\cite{Ford:1992pn}, with 3-loop contributions known at leading order 
in QCD and top Yukawa couplings \cite{Martin:2013gka}, including resummation
of infrared-singular Goldstone boson contributions 
\cite{Martin:2014bca,Elias-Miro:2014pca} (see also 
\cite{Pilaftsis:2015bbs,Kumar:2016ltb,Espinosa:2016uaw,Braathen:2016cqe}
for further developments), 
and at 4-loop order at leading order in QCD \cite{Martin:2015eia}.
Another possible application is to the computation of 
self-energy functions and higher point functions, for which
the results of the present paper can be used in the limit of zero external momentum, 
or in systematic expansions in small external momentum. For example, in supersymmetry,
loop corrections depend on a large number of distinct heavier
superpartner masses. At the present time, the mass hierarchies of the superpartner sector are conjectural,
at best, so that for the foreseeable future it seems most
useful to present results in terms of basis functions 
that can then be evaluated numerically on demand.

{\it Acknowledgments:} We thank Ayres Freitas for 
discussions and comparisons regarding his independent solution 
of the same problem \cite{Freitas}, 
and Mikhail Kalmykov for useful comments.
SPM has also benefited from some long-ago email discussions with 
Oleg Tarasov. This work was supported in part by the National Science 
Foundation grant numbers PHY-1417028 and PHY-1719273.
DGR is supported by a grant from the Ohio Supercomputer Center.

%%%%%%%%%%%%%%%%%%%%%%%%%%%%%

\end{document}